\documentclass[extra,mreferee]{gji}
\usepackage[pdftex]{color,graphicx}
\usepackage{amssymb}
\usepackage[fleqn]{amsmath}
\usepackage{float}
\usepackage{subfig}

\usepackage{times}

\usepackage{color}
\usepackage{booktabs}

\usepackage{multirow}

\usepackage[%
  pdfstartview=FitV,%
  bookmarks=false,%
  bookmarksopen=false,%
  breaklinks=true,%
  colorlinks=true,%
  linkcolor=blue,anchorcolor=blue,%
  citecolor=blue,filecolor=blue,%
  menucolor=blue,%
  urlcolor=blue]{hyperref}
  
\usepackage{rotating}

\usepackage[final]{changes}   
\newcommand{\stkout}[1]{\ifmmode\text{\sout{\ensuremath{#1}}}\else\sout{#1}\fi}
\title[Secular Variation Signals in Magnetic Field Gradient Tensor Elements]{Secular Variation Signals in Magnetic Field Gradient Tensor Elements derived from satellite-based Geomagnetic Virtual Observatories}

\author[Hammer, Finlay and Olsen]{Magnus D. Hammer, Christopher C. Finlay and Nils Olsen\\
Division of Geomagnetism, Technical University of Denmark - DTU Space, Kgs. Lyngby, Denmark,\\
E-mail: magdh@space.dtu.dk \\
}  
\date{Received XXX; in original form XXX}
\pagerange{\pageref{firstpage}--\pageref{lastpage}}
\volume{XXX}
\pubyear{xxxx}

\begin{document}
\maketitle
\vspace{-1.0cm}
\begin{summary}
We present local time series of the magnetic field gradient tensor elements at satellite altitude derived using a Geomagnetic Virtual Observatory (GVO) approach. Gradient element time series are computed in four-monthly bins on an approximately equal-area distributed worldwide network. This enables global investigations of spati\replaced{o}{al}-temporal variations in the gradient tensor elements. Series are derived from data collected by the {\it Swarm} and CHAMP satellite missions, using vector field measurements and their along-track and East-West differences, when available.
We find evidence for a regional Secular Variation impulse (jerk) event in 2017 in the first time derivative  of the gradient tensor elements.  This event is located at low latitudes in the Pacific region. It has a similar profile and amplitude regardless of the adopted data selection criteria and is well fit by an internal potential field. Spherical harmonic models of the internal magnetic field built from the GVO gradient series show lower \replaced{scatter}{noise} in  near-zonal harmonics compared with models built using standard GVO vector field series. The GVO gradient element series are an effective means of compressing the spatio-temporal information gathered by low-Earth orbit satellites on geomagnetic field variations, which may prove useful for core flow inversions and in geodynamo data assimilation studies.
\end{summary}
\section{Introduction}
\label{sec:2}
The main part of the geomagnetic field is generated in the Earth's fluid outer core by a process known as the geodynamo. Knowledge of how this core field varies with space and time provides information on the fluid flow dynamics in the liquid metal outer core. Although the temporal behaviour of the geomagnetic field is well characterized in time series from ground observatories, a global spati\replaced{o}{al}-temporal study is hampered by the uneven distribution of these observatories. Even though low-Earth-orbit (LEO) satellites do provide good global coverage on timescales of weeks and longer, the direct study of the first time derivative of the core field, the secular variation (SV), from satellite measurements is not straightforward\replaced{. This is because}{as} LEO satellites are not geostationary \added{(i.e. do not have the same orbital period as the Earth's rotation, such that their position does not stay fixed as seen from ground), which  leads}\deleted{leading} to \added{an} ambiguity\replaced{, since it is not possible to establish whether an observed field variation is caused by a temporal or spatial change of the field \citep{Olsen_Stolle_2012} }{between spatial and temporal variations}. Spherical harmonic (SH) field models derived from satellite measurements provide an established way of studying the SV field and its time derivative, the secular acceleration (SA), globally. However, such harmonic functions have global support, which means that a\added{n estimate of the} SV \deleted{prediction} at a specific position may be affected by noise from remote locations.

These issues lead \cite{Mandea_Olsen_2006} to introduce the concept of Geomagnetic Virtual Observatories (GVOs) in space, in which satellite magnetic measurements from within a selected region, collected during one month time windows, were used to derive a local monthly mean vector field at the satellite mean altitude. The resulting GVO time series resemble monthly mean series computed using ground observatory magnetic measurements, by providing the magnetic vector field elements at fixed locations. However, since they are based on satellite data, regular sampling in both space and time is possible. \replaced{Thus, the GVO method provides a tool for compressing satellite magnetic field measurements into a dataset that contains time series distributed in a global grid, and which may also be supplied with error estimates \citep{Hammer_etal_2021b}.}{The GVO method provides a means of compressing satellite data into a manageable dataset with global coverage, together with suitable error estimates.} \cite{Olsen_Mandea_2007} used CHAMP measurements to derive GVO vector \replaced{component}{field} time series, and carried out a global investigation of SV that identified a regional geomagnetic jerk event in 2003. 

In the original GVO approach of \replaced{\cite{Mandea_Olsen_2006}}{Mandea and Olsen}, processing of the satellite measurements followed that of simple monthly field means at ground observatories, taking measurements from all local times and with all levels of geomagnetic activity, and relied on the assumption that short period external fields would have zero mean over the course of one month. However, later studies revealed that external fields, especially \added{those} due to the magnetospheric ring current and ionospheric current systems, cause contamination of the retrieved internal GVO field signal  \citep{Olsen_Mandea_2007,Beggan_etal_2009,Domingos_etal_2019}. In addition, insufficient local time sampling from within one month of polar orbiting satellites resulted in a bias due to the local time dependence of ionospheric and magnetosphere-ionosphere coupling currents \citep{Shore_2013}. Recently, the GVO processing algorithm has been further developed in an effort to reduce contamination from magnetospheric and ionospheric sources, and the local time sampling bias, with the aim of better isolating the field signal generated by the Earth's outer core \citep{Hammer_etal_2021a}\deleted{\mbox{\citep{Cox_etal_2020}}}. These new GVO vector field series have been used to study global patterns of field changes \citep{Hammer_etal_2021a,Hammer_etal_2021b}, for inferring fluid flows close to the core surface \citep{Kloss_Finlay_2019,Rogers_etal_2019} and for data assimilation studies \citep{Barrois_etal_2018,Huder_etal_2020}. 

In parallel to the development of these GVO-based techniques there has also been recent progress in the theory of space-based magnetic gradiometry, inspired by advances in satellite gravimetry. Initial studies have demonstrated that knowledge of the second-order $3\times3$ magnetic gradient tensor may be beneficial when seeking to retrieve small scale features of the field (both the lithospheric field and the time-dependent core field). This is possible because gradient elements effectively give more weight to shorter wavelengths, while at the same time some noise sources (e.g. unmodeled magnetospheric fields) are predominantly of long wavelength, which can result in a higher signal-to-noise ratio for short wavelengths compared to using the vector field components \citep{kotsiaros_Olsen_2012,kotsiaros_Olsen_2014}. \added{Information on these shorter spatial length scales is crucial for core dynamic studies, yet their robust determination is a major obstacle in core field modelling. This issue has motivated us to derive time series of field gradient elements using the GVO method, and to investigate whether such series allow, via for example spherical harmonic analysis, an improved retrieval of field changes. Gradient GVO series have the potential to also be used directly in core flow inversions and data assimilation studies of the geodynamo.}

Assuming a potential field due to \deleted{an }internal source\added{s} and no in-situ electrical currents, the field becomes a solenoidal irrotational vector field and the gradient tensor has the special property of being symmetric with a trace of zero. The assumption of a symmetric gradient tensor reduces the number of independent gradient tensor elements from nine to six, while a trace of zero further reduces this number to five. Each element of the magnetic gradient tensor may be considered as a directional filter providing specific information on the magnetic field structures. Thereby, certain gradient tensor elements better constrain specific spherical harmonics \citep{Olsen_Kotsiaros_2011}. According to the studies of \cite{kotsiaros_Olsen_2012} and \cite{kotsiaros_Olsen_2014}, knowledge of the radial gradient of the radial field, written as $\left[\nabla B\right]_{rr}$, is particularly suitable for resolving the higher degree parts and zonal harmonics. The East-West gradient of the azimuthal field, $\left[\nabla B\right]_{\phi \phi}$, and radial field, $\left[\nabla B\right]_{r \phi}$, are especially sensitive towards sectorial harmonics, while the North-South gradient of the radial, $\left[\nabla B\right]_{r \theta}$, and meridional, $\left[\nabla B\right]_{\theta \theta} $, fields are especially useful for determining near-zonal harmonics. The East-West gradient of the meridional field, $\left[\nabla B\right]_{\theta \phi} $ does not provide significant additional information. In addition, knowledge of how certain external fields may influence certain gradient tensor elements is important to consider, for instance the magnetospheric ring current is expected to affect zonal terms constrained by the $\left[\nabla B\right]_{rr}$ element but not the $\left[\nabla B\right]_{r \phi}$ element \citep{kotsiaros_Olsen_2014}. Although it is not yet possible to directly measure the full magnetic gradient tensor in space \citep{Nogueira_etal_2015}, it is nonetheless possible to compute the tensor elements from a magnetic potential determined using satellite magnetic measurements.


In this paper, \replaced{our aim is to}{we} estimate local time series of the magnetic field gradient tensor elements using the GVO method. \added{Our primary motivation for deriving such gradient field series, is to improve the recovery of the small length-scales of the field’s secular variation, compared with the traditional use of vector field data. In particular, the gradient elements are expected to enable a higher signal-to-noise ratio as compared to vector elements \citep{kotsiaros_Olsen_2014}. This is because they are more sensitive to the small length scales of the field, and less affected by large scale external fields.} \replaced{We use satellite magnetic measurements to derive the GVO gradient series, and}{We} follow \cite{Hammer_etal_2021a} in implementing dark and quiet time data selection criteria and use 4-month time windows to minimize problems related to local time sampling\deleted{ \mbox{ \citep{Hammer_etal_2021a}}}. 

In Section \ref{sec:3} we provide a detailed description of the satellite magnetic measurements and selection criteria used, and in Section \ref{sec:4} we describe the GVO method and \added{the} computation of \added{both} GVO \added{vector and GVO} gradient element  time series. In Section \ref{sec:5.1} we present results of the GVO series for each of the SV gradient elements, and visually inspect these. In order to investigate the possible benefits of using GVO gradient data, we compare SH field models derived epoch by epoch from the GVO vector data and GVO gradient tensor data in Section \ref{sec:5.2}. We study the detailed behaviour of the gradient tensor elements going from 2015 to 2018 with \added{a} focus on the Pacific region. Finally, Section \ref{sec:6} provides discussions and conclusions.

\section{Data}
\label{sec:3}
To derive the GVO time series we select vector magnetic field measurements from the CHAMP and {\it Swarm} satellite missions. We used CHAMP L3 magnetic data between July 2000 and September 2010 and {\it Swarm} Level 1b MAG-L, version 0505/0506, from all three {\it Swarm} satellites Alpha, Bravo and Charlie between January 2014 and April 2020, and sub-sample at 15s intervals \added{(i.e. taking measurements every 15 second),} in the vector field magnetometer (VFM) frame. Next, the magnetic data in the VFM frame are rotated into an Earth-Centered Earth-Fixed (ECEF) local Cartesian North-East-Centre (NEC) coordinate frame (for details see \cite{Olsen_etal_2006}) by using the Euler rotation angles from the CHAOS-7.2 model \citep{Finlay_etal_2020}. Measurements from known problematic days (e.g. where satellite manoeuvres took place) were removed and gross data outliers for which the vector field components deviated more than 500nT from CHAOS-7.2 field model \added{[\url{http://www.spacecenter.dk/files/magnetic-models/CHAOS-7/}], see also} \cite{Finlay_etal_2020} \replaced{values}{predictions} were rejected. The measurements were then selected using a dark quiet time criteria defined here as: a) the sun is at least $10^{\circ}$ below horizon, b) geomagnetic activity index $K_p < 3^{\circ}$, c) ring current index $\vert dRC/dt \vert < 3 \mathrm{nT hr^{-1}}$ \citep{Olsen_etal_2014}, merging electric field \added{(averaged over two-hours)} at magnetopause $E_m \leq 0.8 \mathrm{mV m^{-1}}$ \citep{Olsen_etal_2014}, and placing constraints on the interplanetary magnetic field (IMF) \added{(averaged over two-hours)} requiring $B_z > 0\mathrm{nT}$ and $\vert B_y \vert <10 \mathrm{nT}$ \citep{Ritter_etal_2004}. \added{Using the definition in \cite{Olsen_etal_2014}, the merging electric field was derived using 1-min values of the solar wind, solar clock angle and IMF extracted} \deleted{Here we computed two-hourly means of 1 min values of the solar wind and IMF computed} \replaced{from}{form} the OMNI database, \url{http://omniweb.gsfc.nasa.gov}.

Previous studies have demonstrated the benefits of using along-track differences of the satellite magnetic field measurements for retrieving higher spatial resolution of the core and also the lithospheric fields, as such differences filter\deleted{s} out correlated noise caused by external sources \citep{Olsen_etal_2015,Kotsiaros_etal_2015,Kotsiaros_2016,Finlay_2019}. \replaced{However, from data differences alone it is not possible to obtain robust information on the longer wavelength part of the field. Therefore, we also use the complementary data means in a similar manner as in previous studies of the core signal}{To facilitate sufficient constraints on the longer wavelengths of the field, we supplement by include data means} \citep{Sabaka_etal_2013,Hammer_2018}. \replaced{From}{Therefore, from} the satellite magnetic field measurements, $B_k(\mathbf{r})$, where $k$ is the \deleted{unit} vector \added{component} of a given coordinate system, we use measurement means, $\Sigma d_k$, and differences, $\Delta d_k$ as data. The differences, $\Delta d_k=(\Delta d_k^{\mathrm{AT}},\Delta d_k^{\mathrm{EW}})$, and the means, $\Sigma d_k=(\Sigma d_k^{\mathrm{AT}},\Sigma d_k^{\mathrm{EW}})$, are taken along-track (AT) for each satellite and East-West (EW) between the \textit{Swarm} Alpha (SWA) and Charlie (SWC) satellites. Here along-track differences are calculated from the 15~s differences $\Delta d_k^{\mathrm{AT}} = [B_k(\mathbf{r},t) - B_k(\mathbf{r}+\delta \mathbf{r},t+15s)]$ while the means are given by $\Sigma d_k^{AT} = [B_k(\mathbf{r},t) + B_k(\mathbf{r}+\delta \mathbf{r},t+15s)]/2$. The East-West differences were calculated as $\Delta d_k^{\mathrm{EW}} = [B_k^{\mathrm{SWA}}(\mathbf{r}_1,t_1) - B_k^{\mathrm{SWC}}(\mathbf{r}_2,t_2)]$, and the means as $\Sigma d_k^{\mathrm{EW}} = [B_k^{\mathrm{SWA}}(\mathbf{r}_1,t_1) + B_k^{\mathrm{SWC}}(\mathbf{r}_2,t_2)]/2$. Considering a given orbit of \textit{Swarm} Alpha, the corresponding \textit{Swarm} Charlie measurement were chosen to be that closest in colatitude provided that $\vert\Delta t\vert=\vert t_1-t_2\vert<50s$ \citep{Olsen_etal_2015}.

\section{Theory and Method}
\label{sec:4}
\added{The purpose of this paper is to derive time series of the magnetic field gradient tensor elements using the GVO method. In section \ref{sec:4.1}, we begin by recalling the GVO method and how this is used to derive time series of the magnetic field vector elements. Following this, in section \ref{sec:4.2}, we describe our new extension of the GVO method and how this can be used to derive time series of the magnetic field gradient tensor elements.}

\subsection{Geomagnetic Virtual Observatory Method}
\label{sec:4.1}
The Geomagnetic Virtual Observatory method allows for epoch estimates of the magnetic vector field components at a given target point (referred to as a GVO target location) to be derived \replaced{using available satellite measurements enclosed within a cylinder of radius}{satellite measurements from within a region closer than} 700\,km during the course of four months. A radius of 700\,km enables enough data for computing reliable and independent GVO estimates every four months \citep{Hammer_2018}. From these measurements, provided in an ECEF coordinate frame given by the spherical polar components, $\mathbf{B}^{obs}=(B_r,B_{\theta},B_{\phi})$, magnetic field residuals are calculated as \citep{Hammer_etal_2021a}
\begin{equation}
\delta \mathbf{B} = \mathbf{B}^{obs} - \mathbf{B}^{MF} -\mathbf{B}^{lit}- \mathbf{B}^{mag}- \mathbf{B}^{iono} \;,  \label{eq:mag_res}
\end{equation}
where model fields subtracted are: the main field (MF), $\mathbf{B}^{MF}$, for SH degrees $n \in [1,13]$ determined using the CHAOS-7\added{.2} model \added{[\url{http://www.spacecenter.dk/files/magnetic-models/CHAOS-7/}], see also} \cite{Finlay_etal_2020}\replaced{;}{,} the static lithospheric field, $\mathbf{B}^{lit}$, for SH degrees $n \in [14,185]$ determined using the LCS-1 model \citep{Olsen_etal_2017}\replaced{;}{,} the large-scale magnetospheric and associated Earth induced fields, $\mathbf{B}^{mag}$, as given by the CHAOS-7\added{.2} model parameterized in time by the RC index \citep{Finlay_etal_2020}\replaced{;}{,} and the ionospheric and associated Earth induced fields, $\mathbf{B}^{iono}$, as determined using the CIY4 model parameterized by 90-day averages of solar flux F10.7 \citep{Sabaka_etal_2018}. 

Note here that we remove \replaced{values}{predictions} of the \added{time-dependent} main field \replaced{and then at}{in order to facilitate a robust estimation. At} a later stage, \added{add back} main field \replaced{values}{predictions} at the GVO target position and \added{target} epoch \deleted{are added back} \citep{Mandea_Olsen_2006,Hammer_etal_2021a}\replaced{. H}{, h}owever, the precise choice of main field used in both steps is not crucial \citep{Hammer_2018,Hammer_etal_2021b}. \added{Note that in removing estimates of the time-dependent main field in eq.\eqref{eq:mag_res} estimates of the SV field (up to degree 13) are also effectively removed for each data point during the 4 month time windows, which aids a robust estimation. Adding back a main field estimate at the GVO epoch time then synchronizes the data from the considered four month window to this common epoch. This way of correcting the individual data includes spatial gradients of the main field model used, so the resulting local potential does no longer contain the tensor information from the main field model which has been removed. It is thus necessary to reinstate the tensor information from the main field model when computing the magnetic field gradient tensor in section \ref{sec:4.2}.} \added{Note that despite implementing a dark quiet time data selection scheme and also having removed estimates of magnetospheric and ionospheric fields together with their associated Earth-induced fields, contamination from non-core electrical currents may persist in the residual GVO field eq.\eqref{eq:mag_res}. Such contamination could be related to ionospheric currents such as the polar electrojets and F-region currents.}

\replaced{Next, for each data point, both the position of the data point and the corresponding residual magnetic vector, eq.\eqref{eq:mag_res}, are transformed}{Next, the residual magnetic field vector, eq.\eqref{eq:mag_res}, and its positions are transforme } from the spherical system to a right-handed local topocentric Cartesian system $(x,y,z)$ having its origin at the GVO target location, as detailed in \cite[p.~64]{Hammer_2018}. At this specific GVO location (and only at this location), $x$ points towards geographic south, $y$ points towards east and $z$ points radially upwards \citep{Hammer_etal_2021a}. 

At the GVO target point, the unit vectors of the local Cartesian frame \replaced{coincide with the spherical polar unit vectors, i.e. $(\mathbf{\hat{e}_z},\mathbf{\hat{e}_x},\mathbf{\hat{e}_y}) = (\mathbf{\hat{e}_r},\mathbf{\hat{e}_{\theta}},\mathbf{\hat{e}_{\phi}})$}{, $(\mathbf{\hat{e}_x},\mathbf{\hat{e}_y},\mathbf{\hat{e}_z})$, also coincides with the spherical polar unit vectors $(\mathbf{\hat{e}_r},\mathbf{\hat{e}_{\theta}},\mathbf{\hat{e}_{\phi}})$.}. Assuming that the magnetic field measurements are made in a source free region, the residual field, $\delta \mathbf{B}$, is a Laplacian potential field \replaced{that is approximately}{which fulfils the} quasi-stationary \deleted{approximation} \citep{Sabaka_etal_2010}. This means that a magnetic scalar potential, $V$, is associated with the residual field, which in the local Cartesian coordinate system can be expanded as a sum of polynomials having the form $C_{abc} x^a y^b z^c$ \citep{Backus_etal_1996}. In this application we expand to cubic terms following  \cite{Hammer_etal_2021a}
\begin{align}
&V(x,y,z) =  C_{100}x + C_{010}y + C_{001}z + C_{200}x^2 + C_{020}y^2  \nonumber \\
& -(C_{200}+C_{020})z^2  + C_{110}xy + C_{101}xz + C_{011}yz    \nonumber \\
&-\frac{1}{3}(C_{102}+C_{120})x^3 -\frac{1}{3}(C_{210}+C_{012})y^3 \nonumber \\
&-\frac{1}{3}(C_{201}+C_{021})z^3 + C_{210}x^2y+ C_{201}x^2z \nonumber \\  
&+ C_{120}y^2x + C_{021}y^2z + C_{102}z^2x + C_{012}z^2y + C_{111}xyz \; . \label{eq:pot_backus} 
\end{align}
The means and differences of the residual magnetic field are linked to this potential via appropriate design matrices constructed as described in \cite{Hammer_etal_2021a}. The coefficients of the potential are determined from a robust least-squares solution, which includes a) an a prior data covariance matrix derived from standard deviations of the residuals between the data (means and differences) and \replaced{estimates}{predictions} of an un-weighted least-squares solution, b) a diagonal weight matrix consisting of robust (Huber) weights, using a scale constant of 1.5 \citep[e.g.,][]{Constable_1988}, and c) an additional down-weighting factor of 1/2 when data comes from {\it Swarm} satellites Alpha and Charlie, taking into account that these fly side-by-side and thus provide similar measurements. From these potential coefficients, a mean residual magnetic field for the given GVO target point position and epoch is computed as $\delta \mathbf{B}_{GVO}(x\added{=0},y\added{=0},z\added{=0})=-\nabla V=-(C_{100},C_{010},C_{001})$. This mean residual field is then rotated back into the vector components in spherical polar coordinates, $\delta B_{GVO,r}=\delta B_{GVO,z}$, $\delta B_{GVO,\theta}=\delta B_{GVO,x}$, $\delta B_{GVO,\phi}=\delta B_{GVO,y}$ and afterwards a main field \replaced{estimate}{prediction} evaluated at the GVO epoch using SH degrees $n \in [1,13]$ is added back to obtain the GVO field \citep{Hammer_etal_2021a}. \added{We recall here, that the CHAOS-7.2 main field up to degree 13 has to be reinstated, as time-dependent main field estimates were removed by eq.\eqref{eq:mag_res}.}

\begin{figure}
\centerline{\includegraphics[width=1.0\textwidth]{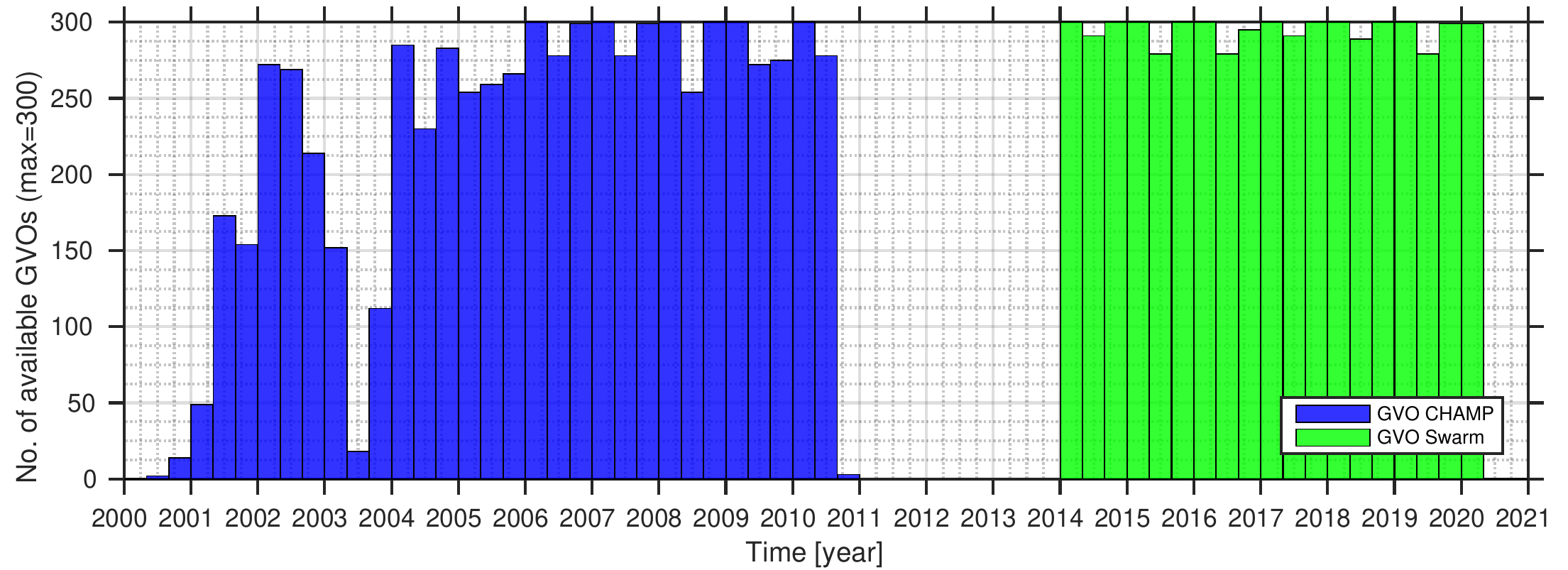}}
\caption{Number \added{(No.)} of \deleted{stable} GVOs for each epoch, given the applied dark and quiet-time selection criteria, during CHAMP (blue) and {\it Swarm} (green) times.}
\label{Fig:2}
\end{figure} 
Following the same procedure as in \cite{Hammer_etal_2021a}, we compute a global grid of 300 GVO's located on an approximately equal area grid based on the sphere partitioning algorithm of \cite{Leopardi_2006}. \added{The global grid of 300 GVOs are listed starting from a position of latitude $89.9^{\circ}$N at longitude $0^{\circ}$ going to a position of latitude $89.9^{\circ}$S at longitude $0^{\circ}$}. The distance between the GVOs in this grid is $\approx1400$\,km, and with a target cylinder radius of 700\,km close to 80\% of the measurements are used. \added{This choice of radius avoids overlap between the GVO data cylinders, such that each satellite measurement is only used once in the global GVO grid computation, which helps to minimize possible correlated errors and biases due to the GVO binning method \citep{Beggan_etal_2009,Shore_2013}.} The GVO height above ground is taken to be 370\,km and 490\,km (approximate mean orbital altitude) during CHAMP and \textit{Swarm} times, respectively. \deleted{The global grid has GVOs located at the North and South poles, where we define the $(r,\theta,\phi)$ frame such that $r$ points radially outwards, $\theta$ is aligned along the Greenwich meridian and $\phi$ completes the right-handed system. In order to add back a main field \replaced{estimate}{prediction} at these two positions, we compute an average value of model field \replaced{values}{predictions} computed $0.1^{\circ}$ in latitude from the North/South Pole at longitudes $0^{\circ}$ and $180^{\circ}$.} Figure \ref{Fig:2} presents the available number of GVOs for each epoch (the maximum possible number at each epoch is 300). \added{Especially when considering the early CHAMP period, there are several epochs with very few available GVOs. Such strongly depleted epochs are primarily caused by our restrictions to geomagnetically quiet and night-time data, and to selection based on CHAMP data quality flags.} Table \ref{table.1} presents the \added{Huber-weighted} mean and root-mean-square (rms) of residuals between the satellite measurements \replaced{(after applying the corrections of eq.\eqref{eq:mag_res}) and the field estimates as computed from the determined potential coefficients $(C_{100},C_{010},C_{001})$. These residuals are}{used for each GVO and the GVO model \replaced{values}{predictions}, and} summed \deleted{up} for each component \added{and} split into regions of 78 polar and 222 non-polar GVO's, defining polar to be GVOs poleward of $\pm 54^{\circ}$ geographic latitude. The polar rms values for both data sums and differences are higher than the non-polar, and the CHAMP values are higher than the \textit{Swarm} values. The non-polar rms values for all components are below 2nT during both CHAMP and {\it Swarm} times. 
\begin{table}
\caption{GVO model rms misfit statistics between contributing satellite data and GVO estimates \replaced{from the total number (No.) of GVOs, as \replaced{compiled}{complied} in Figure \ref{Fig:2},}{using a global grid of 300 GVO's} during CHAMP and \textit{Swarm}. Here {\scriptsize $\sum$} and $\Delta$ represent data means and data differences, respectively.}
\centering
{\small
\begin{tabular}{l r r r r r r | l r r r r r r}
\toprule
           & \multicolumn{3}{c}{CHAMP}  & \multicolumn{3}{c|}{Swarm} & & \multicolumn{3}{c}{CHAMP}  & \multicolumn{3}{c}{Swarm} \\
\cmidrule(lr){2-4} \cmidrule(lr){5-7} \cmidrule(lr){9-11} \cmidrule(lr){12-14}
 Component       & No.    & Mean  & rms  & No.  & Mean  & rms   &  Component          & No.  & Mean  & rms  & No. & Mean  & rms \\
                 &        &  [nT] & [nT] &      &  [nT] &  [nT] &                     &      &  [nT] &  [nT]&     &  [nT] &  [nT] \\
\midrule   
 \textbf{Polar}  & 2574   &       &      & 1872 &        &      & \textbf{Non-polar}& 7326   &       &      & 5328 &        &   \\
 $\sum B_{x,NS}$ &        & -0.30  & 6.61 &     & -0.52  & 6.26   & $\sum B_{x,NS}$ &        &-0.80  & 1.75 &      & 0.01   & 1.69  \\
 $\sum B_{y,NS}$ &        & 0.00  & 6.52 &      & -0.02  & 6.79   & $\sum B_{y,NS}$ &        & 0.00  & 1.46 &      & 0.00   & 1.74\\
 $\sum B_{z,NS}$ &        & 0.00  & 3.33 &      &  0.01  & 3.02   & $\sum B_{z,NS}$ &        & 0.00  & 1.30 &      & -0.00  & 0.95 \\
 $\sum B_{x,EW}$ &        &       &      &      & 0.05   & 5.92   & $\sum B_{x,EW}$ &        &       &      &      & -0.03  & 1.57\\
 $\sum B_{y,EW}$ &        &       &      &      &-0.01   & 6.44   & $\sum B_{y,EW}$ &        &       &      &      & 0.01   & 1.48\\
 $\sum B_{z,EW}$ &        &       &      &      & 0.01   & 2.89   & $\sum B_{z,EW}$ &        &       &      &      &-0.01   & 0.88\\                                             
 $\Delta B_{x,NS}$ &      & -0.01 & 4.35 &      & 0.01   & 3.80   & $\Delta B_{x,NS}$ &      & -0.01 & 0.50 &      & 0.00   & 0.26\\
 $\Delta B_{y,NS}$ &      & -0.01 & 5.20 &      & -0.01  & 4.86   & $\Delta B_{y,NS}$ &      & 0.00  & 0.58 &      & 0.00   & 0.38\\
 $\Delta B_{z,NS}$ &      & 0.01  & 1.61 &      & -0.00  & 1.36   & $\Delta B_{z,NS}$ &      & 0.00  & 0.53 &      & 0.00   & 0.27\\
 $\Delta B_{x,EW}$ &      &       &      &      & 0.10   & 3.17   & $\Delta B_{x,EW}$ &      &       &      &      & 0.10    & 0.51\\
 $\Delta B_{y,EW}$ &      &       &      &      & 0.00   & 3.17   & $\Delta B_{y,EW}$ &      &       &      &      & 0.02  & 0.70\\
 $\Delta B_{z,EW}$ &      &       &      &      & -0.07  & 0.95   & $\Delta B_{z,EW}$ &      &       &      &      & -0.02  & 0.50\\ 
\bottomrule                
\end{tabular}
\label{table.1}}
\end{table}

\subsection{Computing the Magnetic Field Gradient Tensor within the GVO framework}
\label{sec:4.2}
\replaced{W}{In this section w}e now proceed to formulate the magnetic field gradient tensor and describe how \replaced{it}{this} transforms from a spherical polar coordinate system to the local topocentric Cartesian right-handed coordinate system used in the GVO method. This transformation will allow us to compute GVO time series for the magnetic field gradient tensor elements in analogy to the concept of GVO vector field time series. 

We begin by expressing the magnetic field gradient tensor in the local Cartesian system of the GVO method described in Section \ref{sec:4.1}. This is given by (see Appendix \ref{app:A} for full details) 
\begin{align}
\nabla \mathbf{B} &=  - \left(
\begin{array}{lll}
\frac{\partial^2 V}{\partial z^2}          & \frac{\partial^2 V}{\partial x \partial z} & \frac{\partial^2 V}{\partial y \partial z} \\
\frac{\partial^2 V}{\partial z \partial x} & \frac{\partial^2 V}{\partial x^2}          & \frac{\partial^2 V}{\partial y \partial x} \\
\frac{\partial^2 V}{\partial z \partial y} & \frac{\partial^2 V}{\partial x \partial y} & \frac{\partial^2 V}{\partial y^2}
\end{array}\right)= 
\left(
\begin{array}{lll}
\left[\nabla B\right]_{zz}          & \left[\nabla B\right]_{zx}  & \left[\nabla B\right]_{zy} \\
\left[\nabla B\right]_{xz}          & \left[\nabla B\right]_{xx}  & \left[\nabla B\right]_{xy} \\
\left[\nabla B\right]_{yz}          & \left[\nabla B\right]_{yx}  & \left[\nabla B\right]_{yy}
\end{array}\right) \; .  \label{eq:tensor_cartesian_text}
\end{align}
This is a second-order tensor where the minus sign comes from defining the magnetic field as the negative gradient of the potential. The gradient tensor elements are denoted here by $\left[\nabla B \right]_{jk}$, where the first subscript, $j$ denotes the vector component under consideration and the second subscript, $k$, denotes direction of the field derivative. Using the local cubic potential eq.\eqref{eq:pot_backus} estimated from the residual magnetic field as described in Section \ref{sec:4.1}, a second-order residual field gradient tensor at the GVO target point can be derived using eq.\eqref{eq:tensor_cartesian_text} as
\begin{align}
\nabla \delta \mathbf{B}_{GVO} &= \left(
\begin{array}{lll}
2(C_{200}+C_{020})   & -\deleted{2}C_{101}  & -\deleted{2}C_{011}  \\
-\deleted{2}C_{101}            & -2C_{200}  & -\deleted{2}C_{110} \\
-\deleted{2}C_{011}            & -\deleted{2}C_{110} & -2C_{020}
\end{array}\right) \; .  \label{eq:GVO_tensor_cartesian}
\end{align}
Because the magnetic field is a solenoidal vector field, the divergence is zero, such that the trace of the gradient tensor vanishes, i.e. $tr(\nabla \delta \mathbf{B}_{GVO})=2(C_{200}+C_{020})-2C_{200}-2C_{020}=0$, reducing the number of independent elements from 9 to 8. In addition to this, because the field is a Laplacian potential field, the curl of the field vanishes and the number of independent tensor elements reduces to 5; in other words, the magnetic gradient tensor is \added{diagonally} symmetric \replaced{and its}{with} trace \added{is} zero \citep{kotsiaros_Olsen_2012}.
\newline
\newline
Following the GVO algorithm, gradient tensor estimates from a field model then have to be added back at the GVO target point for harmonic degrees $n \leq 13$. \added{The time-dependent CHAOS-7.2 main field up to degree 13 has here to be reinstated, as it was removed by eq.\eqref{eq:mag_res}, in order to synchronize the data to the particular GVO epoch. Removal of the main field estimates results in the local potential in eq.\eqref{eq:pot_backus} also lacking the main field tensor information; this must therefore be added back, in order to obtain the full magnetic field gradient tensor.} To do this, we will need to consider how the gradient tensor elements in spherical polar and Cartesian coordinate systems are related. The magnetic gradient tensor elements as expressed in the spherical coordinate system are given by \citep{Olsen_Kotsiaros_2011,kotsiaros_Olsen_2012}, see also Appendix \ref{app:A}
\begin{align}
\nabla \mathbf{B} &= \left(
\begin{array}{lll}
-\frac{\partial^2 V}{\partial r^2}         & -\frac{1}{r}\frac{\partial^2\partial V}{\partial \theta \partial r}+\frac{1}{r^2}\frac{\partial V}{\partial \theta} & -\frac{1}{r\mathrm{sin}\theta}\frac{\partial^2 V}{\partial \phi \partial r}+\frac{1}{r^2 \mathrm{sin}\theta}\frac{\partial V}{\partial \phi}
 \\
-\added{\frac{1}{r}}\frac{\partial^2 V}{\partial r \partial \theta}+\frac{1}{r^2}\frac{\partial V}{\partial \theta}  & -\frac{1}{r^2}\frac{\partial^2 V}{\partial \theta^2}-\frac{1}{r} \frac{\partial V}{\partial r} & -\frac{1}{r^2 \mathrm{sin}\theta}\frac{\partial^2 V}{\partial \phi \partial \theta}+\frac{\mathrm{cos}\theta}{r^2 \mathrm{sin}^2\theta}\frac{\partial V}{\partial \phi}  \\
-\frac{1}{r \mathrm{sin}\theta} \frac{\partial^2 V}{\partial r \partial \phi}+\frac{1}{r^2 \mathrm{sin}\theta}\frac{\partial V}{\partial \phi}  & -\frac{1}{r^2 \mathrm{sin}\theta}\frac{\partial^2 V}{\partial \theta \partial \phi}+\frac{\mathrm{cos}\theta}{r^2 \mathrm{sin}^2\theta} \frac{\partial V}{\partial \phi} & -\frac{1}{r^2 \mathrm{sin}^2\theta}\frac{\partial^2 V}{\partial \phi^2}-\frac{1}{r}\frac{\partial V}{\partial r}-\frac{\mathrm{cos}\theta}{r^2 \mathrm{sin}^2\theta}\frac{\partial V}{\partial \theta}
\end{array}\right) \nonumber \\
&= \left(
\begin{array}{lll}
\left[\nabla B\right]_{rr}        & \left[\nabla B\right]_{r\theta}       & \left[\nabla B\right]_{r \phi}  \\
\left[\nabla B\right]_{\theta r}  & \left[\nabla B\right]_{\theta \theta} & \left[ \nabla B\right]_{\theta \phi} \\
\left[\nabla B\right]_{\phi r}    & \left[\nabla B\right]_{\phi \theta}   & \left[\nabla B\right]_{\phi \phi} 
\end{array}\right) \;. \label{eq:tensor_spherical}
\end{align} 
Here the first column of the tensor contains the derivatives of the magnetic field components along the radial direction, the second column contains the derivatives along the co-latitudinal direction and the third column contains the derivatives along the longitudinal direction. The gradient element in the first column and row contains one term only, the field derivative term e.g. $\partial^2/\partial r^2$, while the \replaced{other}{rest of the} gradient tensor elements in addition to this also have an additional field term i.e. $\partial/\partial r$, $\partial/\partial \theta$ or $\partial/\partial \phi$. \deleted{Appendix \ref{app:B} provides example plots of the SV gradient tensor elements at the Earth's surface in 2018, decomposed into the field derivative term, the field term parts and both terms together as computed using the CHAOS-7 field model \mbox{\citep{Finlay_etal_2020}} }\deleted{.} The transformations relating the gradient tensor elements in the local Cartesian system to the tensor elements of the spherical system,  {\it only at the GVO target location}, are in the end given by the following simple relations (see Appendix \ref{app:A} for a full derivation).
\begin{align}
\left[\nabla B \right]_{zz}&= \left[\nabla B \right]_{rr} \qquad
\left[\nabla B \right]_{zx}=\left[\nabla B \right]_{r \theta} \qquad
\left[\nabla B \right]_{zy}=\left[\nabla B \right]_{r \phi}\\ \nonumber
\left[\nabla B \right]_{xz}&=\left[\nabla B \right]_{\theta r}\qquad
\left[\nabla B \right]_{xx}=\left[\nabla B \right]_{\theta \theta}\qquad
\left[\nabla B \right]_{xy}=\left[\nabla B \right]_{\theta \phi}\\ \nonumber
\left[\nabla B \right]_{yz}&=\left[\nabla B \right]_{\phi r} \qquad
\left[\nabla B \right]_{yx}=\left[\nabla B \right]_{\phi \theta} \qquad
\left[\nabla B \right]_{yy}=\left[\nabla B \right]_{\phi \phi} \;. \nonumber
\end{align}
Having determined the potential from the residual magnetic field eq.\eqref{eq:mag_res}, we can compute a residual field gradient tensor by eq.\eqref{eq:GVO_tensor_cartesian} and add back main field gradient tensor estimates from the CHAOS-7.2 field model using eq.\eqref{eq:tensor_spherical} for SH degrees $n \leq 13$, using the above relations, in order to obtain the required GVO field gradient estimates $\nabla \mathbf{B}_{GVO}$. Note that this procedure is analogous to the procedure applied in deriving vector field GVOs where the main vector field is added back. The above procedure is then repeated at each GVO location and for each epoch to compute all the desired GVO field gradient time series. 

\added{Field contamination from non-core sources dominates the instrumental errors of the satellites. Thus, the primary limitation in obtaining core field GVOs are contributions due to unmodelled fields, such as the polar electrojet, and not measurement errors \citep[e.g.][]{Finlay_etal_2017}.} Error estimates for each tensor element $jk$, and separately for CHAMP and \textit{Swarm}, are computed using the residuals $e_{jk}=d^{GVO}-d^{CHAOS}$\deleted{,} between the GVO gradient tensor data, $d^{GVO}=\left[ \nabla B_{GVO} \right]_{jk}$, and the gradient element \replaced{values}{predictions} of the CHAOS-7\added{.2} for SH degree $n=1-16$, $d^{CHAOS}=\left[\nabla B \right]_{jk}$. Considering all epochs for each GVO in the grid, the error estimates for tensor element $jk$ are given by the total mean square error $\sigma_{jk} = \sqrt{\sum_i(e_{jk,i}-{\mu_{jk}})^2 / M + \mu_{jk}^2 }$ \citep[e.g.][]{Bendat_Piersol_2010}, where $e_{jk,i}$ is the residual of the $i$th data element, $M$ is the number of data in a given series and $\mu_{jk}$ is the residual mean for a given component. \cite{Hammer_etal_2021a} computed similar uncertainty estimates for the vector components using the vector field residuals \replaced{with respect to}{towards} the CHAOS-7\added{.2} model. \added{We note here, that using the CHAOS model to quantify the variability of the GVO series, provides only a crude indication of the data errors, which are assumed time independent. Moreover, correlation in the data errors are presently not accounted for, although this is expected especially at high latitudes. It is however challenging to empirically estimate non-diagonal data error covariance matrices with the short GVO time series presently available. Further work is needed on this important topic.}

\begin{table}
\caption{Mean of the rms differences (in $\mathrm{pT/km \: yr^{-1}}$) between GVO SV series and GCV cubic spline fits for six of the gradient tensor elements. Results are shown for GVO SV gradient series derived from \textit{Swarm} and CHAMP data using CHAOS-7.2 \citep{Finlay_etal_2020} as MF model in the GVO processing.}
\centering
{\small
\begin{tabular}{l | c c c c c c}
\hline
 Component          & $rms_{[rr]}$& $rms_{[\theta \theta]}$& $rms_{[\phi \phi]}$& $rms_{[r \theta]}$& $rms_{[r \phi]}$& $rms_{[\theta \phi]}$  \\
                    & $[\mathrm{pT}/\mathrm{km} \: \mathrm{yr}^{-1}]$& $[\mathrm{pT}/\mathrm{km} \: \mathrm{yr}^{-1}]$  & $[\mathrm{pT}/\mathrm{km} \: \mathrm{yr}^{-1}]$ & $[\mathrm{pT}/\mathrm{km} \: \mathrm{yr}^{-1}]$ & $[\mathrm{pT}/\mathrm{km} \: \mathrm{yr}^{-1}]$ & $[\mathrm{pT}/\mathrm{km} \: \mathrm{yr}^{-1}]$    \\
\hline                             
CHAMP               &          &          &          &          &          &         \\ 
\textit{Polar}      &     5.40 &     3.90 &     4.20 &     2.10 &     2.60 &     2.60 \\ 
\textit{non-Polar}  &     2.10 &     0.70 &     2.00 &     0.70 &     1.60 &     0.60 \\ 
\textit{All}        &     2.98 &     1.55 &     2.55 &     1.03 &     1.83 &     1.14 \\  
\hline          
\textit{Swarm}      &          &          &          &          &          &         \\ 
\textit{Polar}      &     4.20 &     3.20 &     3.90 &     1.60 &     1.70 &     2.10 \\ 
\textit{non-Polar}  &     1.20 &     0.30 &     1.20 &     0.30 &     0.70 &     0.60 \\ 
\textit{All}        &     1.98 &     1.01 &     1.86 &     0.61 &     0.95 &     0.97 \\ 
\hline   
\end{tabular}
\label{table:1a}}
\end{table} 

As with the ordinary GVO vector field time series, we estimate GVO gradient tensor time series \replaced{on}{in} a global grid of 300 GVOs. We compute the SV as annual differences at each GVO for each tensor element. In order to quantify the scatter level\deleted{s} in each series, we then fit cubic smoothing splines to the time series, with a knot spacing of 4 months and a smoothing parameter determined using a generalized cross-validation (GCV) approach \citep{Green_Silverman_1993}. Table \ref{table:1a} presents the mean rms differences between the GVO SV gradient tensor elements and GCV spline fits, separated into polar and non-polar regions. These rms numbers provide an indication of the scatter level in the GVO SV gradient data derived from the CHAMP and {\it Swarm} measurements first using CHAOS-\added{7.2} \citep{Finlay_etal_2020} as a main field model. Comparing the numbers between CHAMP and {\it Swarm}, we see that overall the values are lower for {\it Swarm}, i.e. {\it Swarm} gradient tensor element SV time series have a lower scatter than similar series for CHAMP. \replaced{We note that the misfits are considerably lower at non-Polar GVOs, compared with Polar GVOs, and this applies to all of the gradient elements, during both CHAMP and {\it Swarm} times. Furthermore, the off-diagonal elements and the $d\left[\nabla B\right]_{\theta \theta}/dt$ element show considerably lower misfit values being below $2 \: \mathrm{pT/km \: yr^{-1}}$ and $1 \: \mathrm{pT/km \: yr^{-1}}$ at non-Polar latitudes during CHAMP and {\it Swarm} times, respectively.}{In particular, we note that the $d\left[\nabla B\right]_{\theta \theta}/dt$ and $d\left[\nabla B\right]_{r \theta}/dt$ elements show considerably lower misfit values having non-polar values of $0.3 \: \mathrm{pT/km \: yr^{-1}}$ and $0.5 \: \mathrm{pT/km \: yr^{-1}}$, respectively, during {\it Swarm} and $0.7 \: \mathrm{pT/km \: yr^{-1}}$ and $1.4 \: \mathrm{pT/km \: yr^{-1}}$ during CHAMP times, respectively. }

\begin{table}
\caption{Mean of the rms differences (in $\mathrm{pT/km \: yr^{-1}}$) between GVO SV series and GCV cubic spline fits for six of the gradient tensor elements. Results are shown for GVO SV gradient data derived from \textit{Swarm} and CHAMP data using COV-OBS.x2 model \citep{Huder_etal_2020} as MF model in the GVO processing.}
\centering
{\small
\begin{tabular}{l | c c c c c c}
\hline
 Component          & $rms_{[rr]}$& $rms_{[\theta \theta]}$& $rms_{[\phi \phi]}$& $rms_{[r \theta]}$& $rms_{[r \phi]}$& $rms_{[\theta \phi]}$  \\
                    & $[\mathrm{pT}/\mathrm{km} \: \mathrm{yr}^{-1}]$& $[\mathrm{pT}/\mathrm{km} \: \mathrm{yr}^{-1}]$  & $[\mathrm{pT}/\mathrm{km} \: \mathrm{yr}^{-1}]$ & $[\mathrm{pT}/\mathrm{km} \: \mathrm{yr}^{-1}]$ & $[\mathrm{pT}/\mathrm{km} \: \mathrm{yr}^{-1}]$ & $[\mathrm{pT}/\mathrm{km} \: \mathrm{yr}^{-1}]$    \\
\hline                             
CHAMP                &          &          &          &          &          &         \\ 
\textit{Polar}     &     5.40 &     3.90 &     4.20 &     2.10 &     2.60 &     2.60 \\ 
\textit{non-Polar} &     2.10 &     0.70 &     2.00 &     0.70 &     1.60 &     0.60 \\ 
\textit{All}       &     2.99 &     1.55 &     2.55 &     1.03 &     1.83 &     1.14 \\ 
\hline          
\textit{Swarm}      &          &          &          &          &          &         \\ 
\textit{Polar}     &     4.20 &     3.20 &     3.80 &     1.60 &     1.70 &     2.10 \\ 
\textit{non-Polar} &     1.20 &     0.30 &     1.20 &     0.30 &     0.70 &     0.60 \\ 
\textit{All}       &     1.97 &     1.01 &     1.86 &     0.61 &     0.96 &     0.97 \\ 
\hline  
\end{tabular}
\label{table:1b}}
\end{table} 

We also tested how the choice of main field model (used for subtracting and adding back main field estimates) would impact the results. We produced test GVO tensor element series from both CHAMP and {\it Swarm} measurements using the main field \replaced{values}{predictions} for SH degrees $n \in [1,13]$ of the COV-OBS.x2 model \citep{Huder_etal_2020}. Table \ref{table:1b} presents the mean rms differences using the COV-OBS.x2 model instead of CHAOS-7\added{.2}. This results in almost identical misfit levels to the GCV splines (i.e. scatter), between the GVO gradient series during CHAMP and {\it Swarm} times, regardless of whether CHAOS-7.2 or COV-OBS.x2 is chosen.

\section{Results}
\label{sec:5}
\subsection{Field Gradient Element SV time series}
\label{sec:5.1}
\begin{figure}
\centerline{\includegraphics[width=1.0\textwidth]{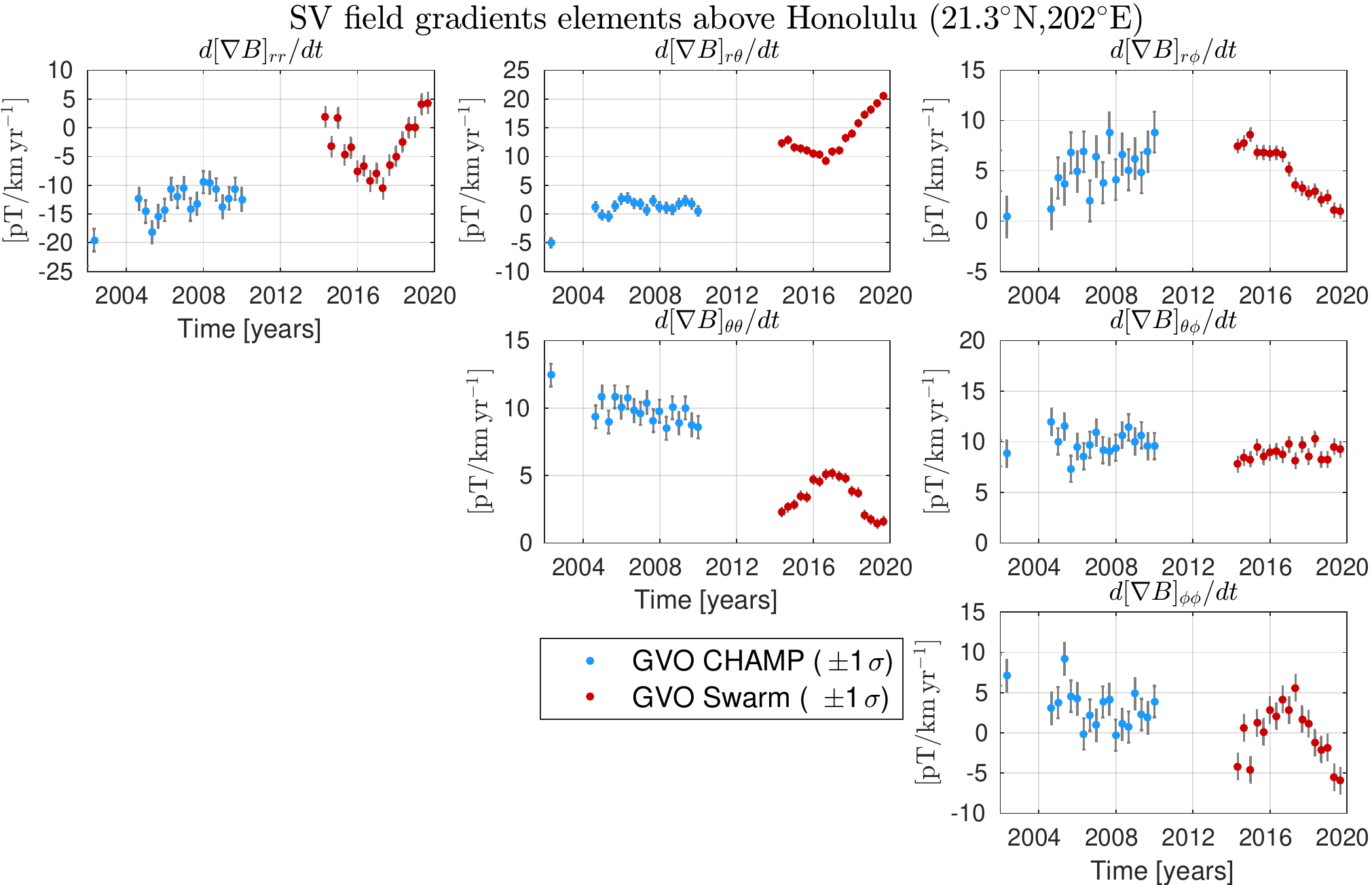}}
\caption{Annual differences of the GVO field gradient elements with $\pm1 \sigma$ uncertainties, during CHAMP (blue) and {\it Swarm} (red) times at altitude 500km for a case study above Honolulu, Hawaii. Units are $\mathrm{pT}/\mathrm{km} \, \mathrm{yr}^{-1}$.}
\label{Fig:7b}
\end{figure}

We begin by investigating the temporal behaviour of the annual differences of each gradient tensor element at an example GVO location above Honolulu ground observatory in Hawaii, from which there are well known vector field records. To do this, we compute dedicated GVO gradient element series above the Honolulu observatory using the method described in Section \ref{sec:4.2}. Here we are motivated by studies which have point\added{ed} out a change in secular acceleration of the radial component in the Pacific occurring around 2017 \citep{Sabaka_etal_2018,Finlay_etal_2020}. In particular, we are interested to see if it is possible to identify this event in the GVO gradient tensor time series, and how this will display in the various tensor elements. Figure \ref{Fig:7b} present plots of the SV for each gradient tensor element above Honolulu, showing the GVOs derived from CHAMP (in blue) and {\it Swarm} (in red) measurements. For comparison purposes we have mapped the two GVO series to a common altitude of 500\,km by subtracting \deleted{off} the SV gradient field differences between the GVO altitudes and 500\,km altitude using the CHAOS-7.2 model. \added{The uncertainty estimates $\pm \sigma$ are computed as for the global grid GVOs, i.e. the total mean square error from residuals between the Honolulu GVO gradient tensor data and gradient estimates from the CHAOS-7.2 model up to degree 16.}

\deleted{We begin by noting that the SV gradient tensor in Figure \ref{Fig:7b} is symmetric, as expected. }Visual inspection clearly demonstrates that geophysical signals are captured in all of the SV gradient tensor elements. Distinctive changes centred around 2017 can be observed having a "$V$" shape in the $d\left[\nabla B\right]_{rr}/dt$ and $d\left[\nabla B\right]_{r \theta}/dt$ elements, with a corresponding "$\Lambda$" shape in the $d\left[\nabla B\right]_{\theta \theta}/dt$ and $d\left[\nabla B\right]_{\phi \phi}/dt$ elements. In addition to this, we note that during 2004-2010, especially the $d\left[\nabla B\right]_{r \theta}/dt$ element displays a variation pattern which resembles that found in the $\theta$-component of the annual differences of monthly mean vector field series from Honolulu (not shown).  
 
\begin{figure}
\centerline{\includegraphics[width=1.0\textwidth]{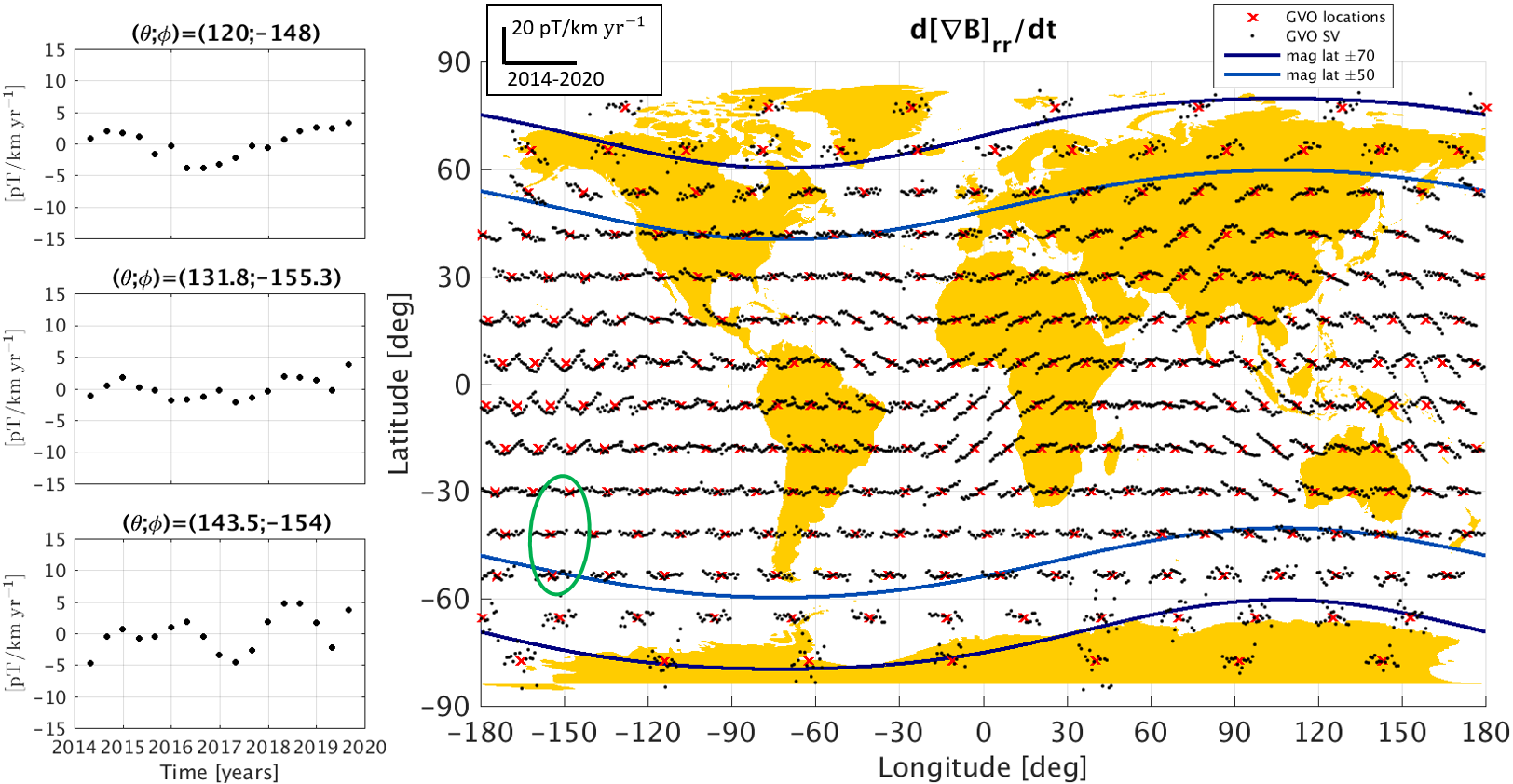}}
\caption{Time series of SV field gradient element $d\left[\nabla B\right]_{rr}/dt$ (map) during {\it Swarm} time from 2014-2020 at 490km altitude. Magnetic latitudes $\pm50^{\circ}$ and $\pm70^{\circ}$ shown with blue curves. GVO locations marked with a red cross. Highlighted are selected time series (locations marked with green ellipse) after removing the mean trend in order to ease comparison, at three GVO locations along the $150^{\circ}$W meridian line at latitudes $30^{\circ}$S (top), $42^{\circ}$S (center) and $54^{\circ}$S (bottom). }
\label{Fig:10b}
\end{figure}

Next, we investigate the global behaviour of annual differences of the gradient elements for GVOs derived from {\it Swarm} measurements during 2014-2020. Here we have chosen to present global series for the $d \left[\nabla B\right]_{rr}/dt$ element in Figure \ref{Fig:10b}. By visual inspection, we find local regions with similar temporal changes as those observed at the Honolulu SV gradient series. In particular, a distinct "$V$" shaped behaviour is found in the eastern Pacific region in a band stretching from latitudes $20^{\circ}$S to $20^{\circ}$N and longitudes $180^{\circ}$ to $220^{\circ}$ with a possibly related opposite "$\Lambda$" shaped behaviour in the western Pacific region from latitudes $20^{\circ}$S to $20^{\circ}$N and longitudes $120^{\circ}$E to $180^{\circ}$E. These regional changes occur over a time window of 6 years reaching amplitudes of about $15 \: \mathrm{pT}/\mathrm{km} \, \mathrm{yr}^{-1}$. Note that a "$V$"-shaped SV gradient time series means a strong positive change in the SA, while a "$\Lambda$"-shaped time series means a strong negative change in the SA. Though more complex to interpret, the other SV gradient tensor elements (not shown) also exhibit distinctive behaviour in the Pacific region. These observed changes in the SV gradient elements indicate regional jerk-type event happening in the Pacific centred on 2017. In the ionosphere external fields tend to be organized according to the geometry of Earth's main field, and their signal in the GVO series may therefore be grouped accordingly to magnetic latitude in quasi-dipole coordinates \citep{Laundal_Richmond_2017}. Here magnetic latitude $\pm70^{\circ}$ (dark blue curve) may be used to approximate the border between North/South Polar and Auroral zones, while magnetic latitude $\pm50^{\circ}$ (light blue curve) divides the North/South Auroral and Low- to Mid-latitude zones \citep{Hammer_etal_2021a}. In all of the SV gradient element maps, higher scatter are found at GVOs located in the Polar and Auroral zones, which is consistent with noise (unmodeled fields) from ionospheric and magnetosphere-ionosphere coupling currents. \added{In addition, some rapid variations can be seen, which decrease in amplitude when going towards equatorial latitudes. An example of this is found along longitude $150^{\circ}$W, stretching from latitudes $30^{\circ}$S to $60^{\circ}$S, as highlighted in the side-panels of Figure \ref{Fig:10b} at three selected GVO locations (marked in the global maps by the green ellipse) }

\deleted{Besides the aforementioned variations, rapid small amplitude SV fluctuations within a few years can be seen especially clear in the $d\left[\nabla B\right]_{rr}/dt$ element. Are such rapid changes of external origin? In particular, two types of variations in the SV gradient series could be indicative of external field leakage: 1) a temporal feature seen in polar latitude GVO time series persisting in series at lower latitudes along the same meridional line, could be an indicator of a contaminating signal of ionospheric or field-aligned current origin, due to the incomplete sampling of local times in the contributing satellite data 2) temporal features seen at mid or low latitudes in the GVO series at all longitudes could be a sign of a signal having magnetospheric origin. In the global time series no distinct similar temporal feature can be observed along all longitudes at mid/low latitudes, thus suggesting that magnetospheric disturbances are small. However, some contamination from ionospheric currents at higher latitudes persists to lower latitudes. Considering for instance $d\left[\nabla B\right]_{rr}/dt$ series along longitude $150^{\circ}$W, stretching from latitudes $30^{\circ}$S to $60^{\circ}$S, some rapid variations can be seen that decrease in amplitude going towards equatorial latitudes, as highlighted in the side-panels of Figure \ref{Fig:10b} at three selected GVO locations (marked in the global maps by the green ellipse).}

\begin{figure}
\centerline{\includegraphics[width=1.0\textwidth]{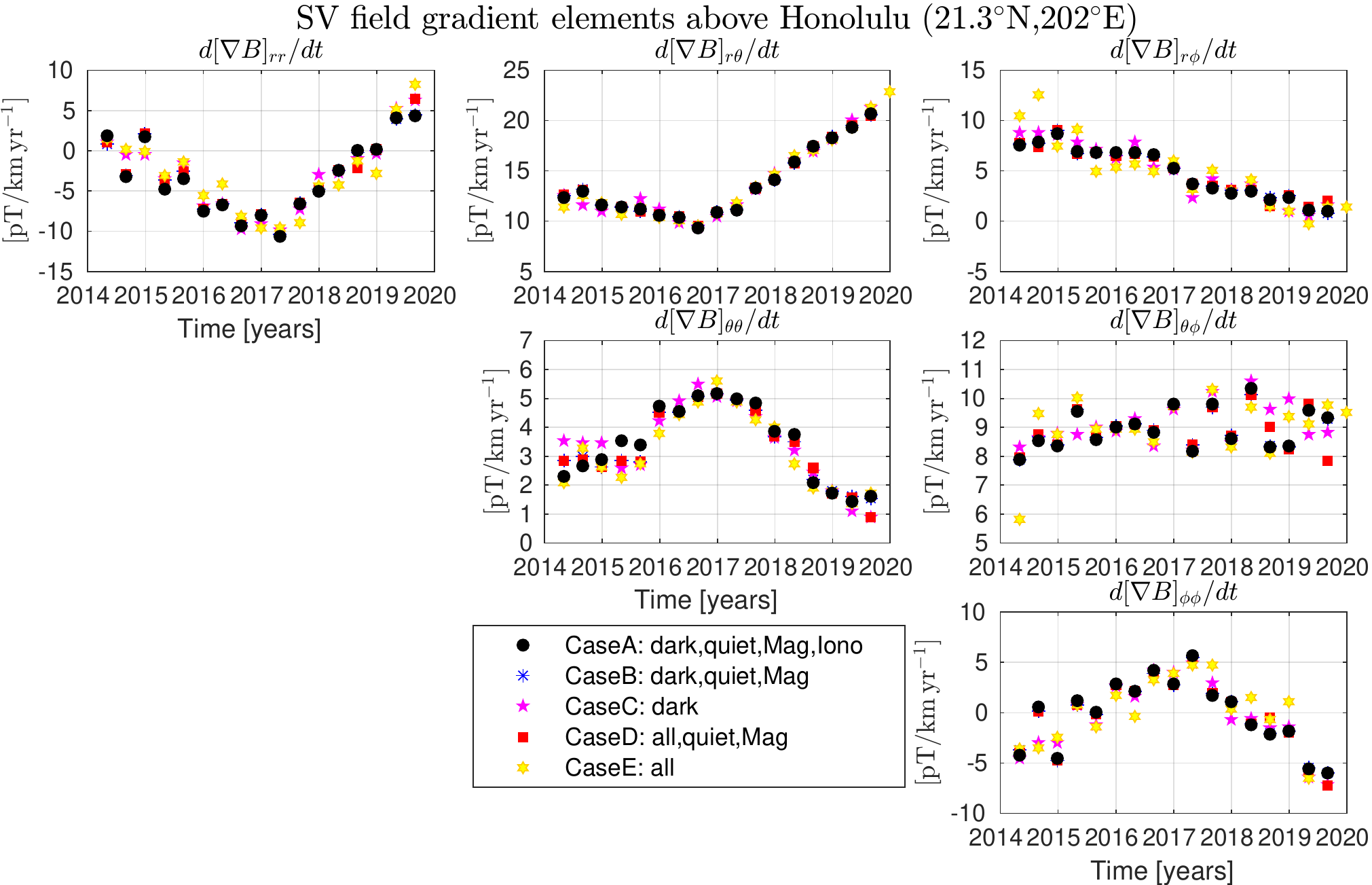}}
\caption{Annual differences of the field gradient elements from GVO's derived using different data selection criteria, as described in the text, during {\it Swarm} time for a case study above Honolulu, Hawaii. Units are $\mathrm{pT}/\mathrm{km} \, \mathrm{yr}^{-1}$.}
\label{Fig:11}
\end{figure} 
An important question is whether the prominent change in SV observed centred on 2017 is robust and of internal origin. To address this, we produced a set of the GVO SV series above Honolulu\deleted{,} during {\it Swarm} time, testing a range of geomagnetic selection criteria. We considered five cases: Case A using a dark quiet time data selection removing estimates of the magnetospheric and ionospheric fields as described in Section \ref{sec:3}, this is our 'preferred' criteria for studying core field variations, Case B using a dark quiet time data selection removing estimates of the magnetospheric field but not removing estimates of the ionospheric field, Case C using a dark time data selection but with neither ionospheric nor magnetospheric corrections, Case D using a quiet time data selection from both day and night ("all" local times) and \replaced{removal}{estimates} of the magnetospheric field\replaced{,}{were removed} and Case E using data from "all" local times, without any quiet time data selection applied and without \replaced{removal of}{corrections} for magnetospheric or ionospheric fields. Here "dark" means the sun is required to be at least $10^{\circ}$ below horizon, and "all" means that no such \replaced{selection is applied}{requirement is used}, i.e. sunlit data are also included. SV gradient series for all five data selection cases are shown in Figure \ref{Fig:11}. The black dots corresponding to Case A are for our "preferred" criteria were also used to derive the maps in Figure \ref{Fig:10b}. All \deleted{the }selection criteria results in the same overall temporal "$\Lambda / V$" shape behaviour with an amplitude of $\approx 15 \: \mathrm{pT/km \: yr^{-1}}$. There is no increase in amplitude \replaced{when}{on} including more disturbed data. For example, comparing Case C (purple star) with Case A/Case B (black/blue dots) should expose a signal from a magnetospheric source; however, the same "$\Lambda / V$" shape behaviour appears in all three cases. Comparing instead Case E (yellow star) with Case C and Case D should expose an ionospheric signal, which is expected to be larger during sunlit conditions; even though more scatter is seen in Case E, the same overall "$\Lambda / V$" shape is clearly visible and \replaced{of}{with} similar amplitude. These results are consistent with an internal origin \replaced{of}{for} the 2017 SV impulse event.

\subsection{Example Spherical Harmonic Models Derived From Gradient Data}
\label{sec:5.2}
In this section we demonstrate that spherical harmonic (SH) field models with high temporal resolution (4 months) can be built from the global network of GVO gradient tensor time series. We then use these models to investigate global change\added{s} in SA during {\it Swarm} time, and in particular, we analyse the possible benefits of the GVO field gradient tensor series over more standard GVO vector field series.
\added{The magnetic field can be described as the negative gradient of internal, $V^{int}$, and external, $V^{ext}$, potentials} \citep[e.g.][]{Sabaka_etal_2010} 
\begin{equation}
\added{\mathbf{B} = -\nabla V^{int} - \nabla V^{ext} \;.} \label{eq:POT_int_ext}     
\end{equation}
\added{Here the internal and external potentials at a given epoch can be expressed by truncated spherical harmonic expansions up to degree $N^{int}$ and $N^{ext}$, respectively}
\begin{align}
\added{V^{int}(r,\theta,\phi,t) }  &\added{= r_a \sum_{n=1}^{N^{int}} \sum_{m=0}^{n} \left[ g_n^m(t) \mathrm{cos}(m\phi) +h_n^m(t) \mathrm{sin}(m\phi) \right] \left(\frac{r_a}{r}\right)^{n+1} P_n^m(\mathrm{cos}\,\theta) }  \label{eq:V_int} \\
\added{V^{ext}(r,\theta,\phi,t)}  &\added{= r_a \sum_{n=1}^{N^{ext}} \sum_{m=0}^{n} \left[ q_n^m(t) \mathrm{cos}(m\phi) +s_n^m(t) \mathrm{sin}(m\phi) \right] \left(\frac{r}{r_a}\right)^{n} P_n^m(\mathrm{cos}\,\theta) } \;,  \label{eq:V_ext}  
\end{align}
\added{where $r_a=6371.2$km is the Earth's mean spherical radius, $n$ and $m$ are the SH degree and order, respectively, and $P_n^m$ are the associated Schmidt semi-normalized Legendre functions. 
$\{g_n^{m},h_n^{m}\}$ and $\{q_n^{m},s_n^{m}\}$ are the internal and external expansion coefficients, respectively.} \replaced{Assuming an internal field source, t}{T}he linear forward problem of determining the SH expansion coefficients can be written

\begin{equation}
\mathbf{d} = \underline{\underline{\mathbf{G}}} \mathbf{m}, \label{eq:10}
\end{equation}

where $\mathbf{d}$ is a data vector containing the GVO epoch data (i.e. field vector components or field gradient tensor elements), $\underline{\underline{\mathbf{G}}}$ is the design matrix for an internal potential relating each model coefficient to \replaced{either the GVO epoch vector or gradient}{the} data \added{(for explicit expressions of the gradient tensor elements, we refer to Appendix A of \cite{Kotsiaros_2012})}, and vector $\mathbf{m}$ contains the parameters of the potential, i.e. the internal SH coefficients here denoted as $g_n^m$ and $h_n^m$ for order $m$ and degree $n$. For each GVO epoch we estimate a \added{single epoch} SH model using \replaced{an Iteratively Reweighted Least Squares method which is applied in order to mitigate the impact of non-Gaussian distributed data residuals \citep[e.g.][]{Olsen_etal_2006}}{a robust least-squares solution}
\begin{equation}
\mathbf{m} = 
(\underline{\underline{\mathbf{G}}}^T \underline{\underline{\mathbf{W}}} \ \underline{\underline{\mathbf{G}}})^{-1}\underline{\underline{\mathbf{G}}}^T \ \underline{\underline{\mathbf{W}}}\mathbf{d} \;, \label{eq:11}
\end{equation}
where \added{we assign weights according to the diagonal weight matrix} $\underline{\underline{\mathbf{W}}}=w/\sigma^2$ \deleted{is a diagonal weighting matrix} consisting of Huber weights, $w$, having a Huber tuning \added{constant} of 1.5 \citep[e.g.,][]{Constable_1988} and error estimates, $\sigma^2$, of either the field vector components or field gradient tensor elements\added{, depending on which type of data is solved for.} \added{These error estimates are derived as the total mean squared error \citep{Bendat_Piersol_2010} of the residuals between the GVO vector/gradient tensor series and vector/gradient estimates of the CHAOS-7.2 model up to degree 16 (see Section \ref{sec:4.2}). That is, the weight matrix consists of the inverse data covariance matrix elements multiplied by Huber weights applied in order to account for non-Gaussian data residuals. The weights are then iteratively updated until convergence.} We derive models up to SH degree \added{$N^{int}=14$} for each 4 month interval. \deleted{No spatial or temporal regularization is applied.} 

\begin{figure}
\centerline{\includegraphics[width=1.0\textwidth]{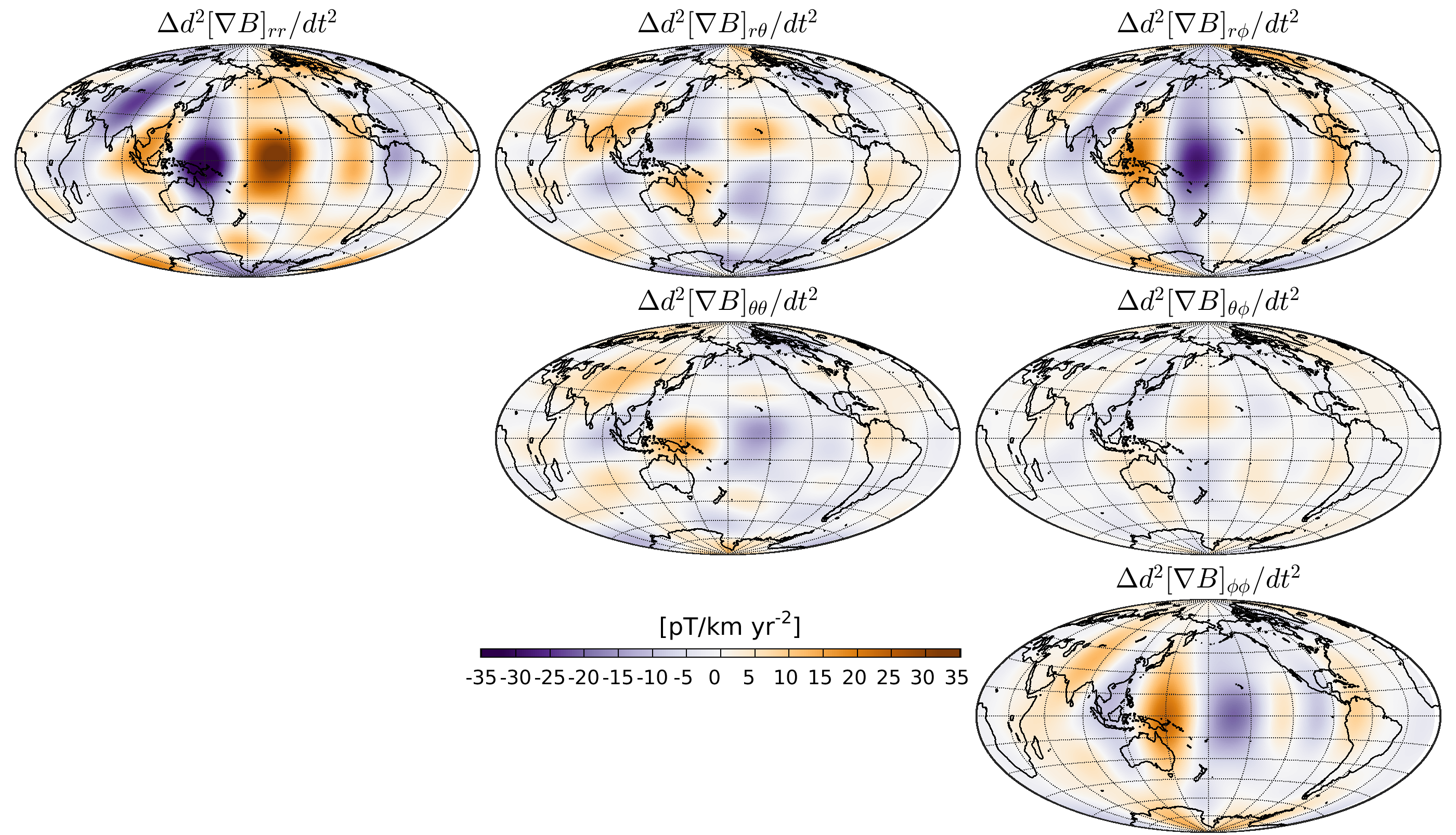}}
\caption{Change in SA gradient tensor elements between 2015.5 and 2018.5 for SH degrees $n\leq9$ at the Earth's surface.}
\label{Fig:13}
\end{figure}

To investigate the 2017 region jerk event we next compute the secular acceleration change for each gradient tensor element between 2015.5 and 2018.5 at the Earth's surface
\begin{equation}
\Delta d^2  \left[\nabla B\right]_{jk}  /dt^2 = d^2 \left[\nabla B\right]_{jk} /dt^2 \vert_{2018.5} - d^2  \left[\nabla B\right]_{jk} /dt^2 \vert_{2015.5} \; . \label{eq:12}
\end{equation}
Plotting global maps of this change in Figure \ref{Fig:13} for each of the gradient elements for degrees $n\leq9$ at the Earth's surface, distinct patterns of SA change are seen to have occurred in the Pacific region during 2015.5-2018.5. Only results for the upper right part of the gradient \replaced{tensor}{elements} are shown as the tensor is symmetric.  The $\Delta d^2 \left[\nabla B\right]_{rr} /dt^2$ map identifies two strong localized patches of opposite sign in SA change reaching amplitudes of $40 \: \mathrm{pT/km \, yr^{-2}}$ in a region \replaced{confined to}{defined by} latitudes $25^{\circ}$S to $25^{\circ}$N and longitudes $140^{\circ}$ to $220^{\circ}$. Associated strong negative and positive patches are seen in the $\Delta d^2 \left[\nabla B\right]_{\phi \phi}  /dt^2$ map in the same region. In addition, the $\Delta d^2 \left[\nabla B\right]_{r \theta} /dt^2$ and $\Delta d^2 \left[\nabla B\right]_{\theta \phi} /dt^2$ elements show a tiling pattern of positive and negative field patches from latitudes $25^{\circ}$S to $25^{\circ}$N and longitudes $120^{\circ}$ to $240^{\circ}$. Similar changes, but in the radial field SA between 2014 to 2020, involving nearby features with opposite sign in the Pacific region, have been found in the CHAOS-7\added{.2} field model \cite[]{Finlay_etal_2020} and using the technique of Subtractive Optimized Local Averages (SOLA) applied to \textit{Swarm} data \cite[]{Hammer_Finlay_2019, Hammer_etal_2021b}. 

\begin{figure}
\centerline{\includegraphics[width=0.49\textwidth]{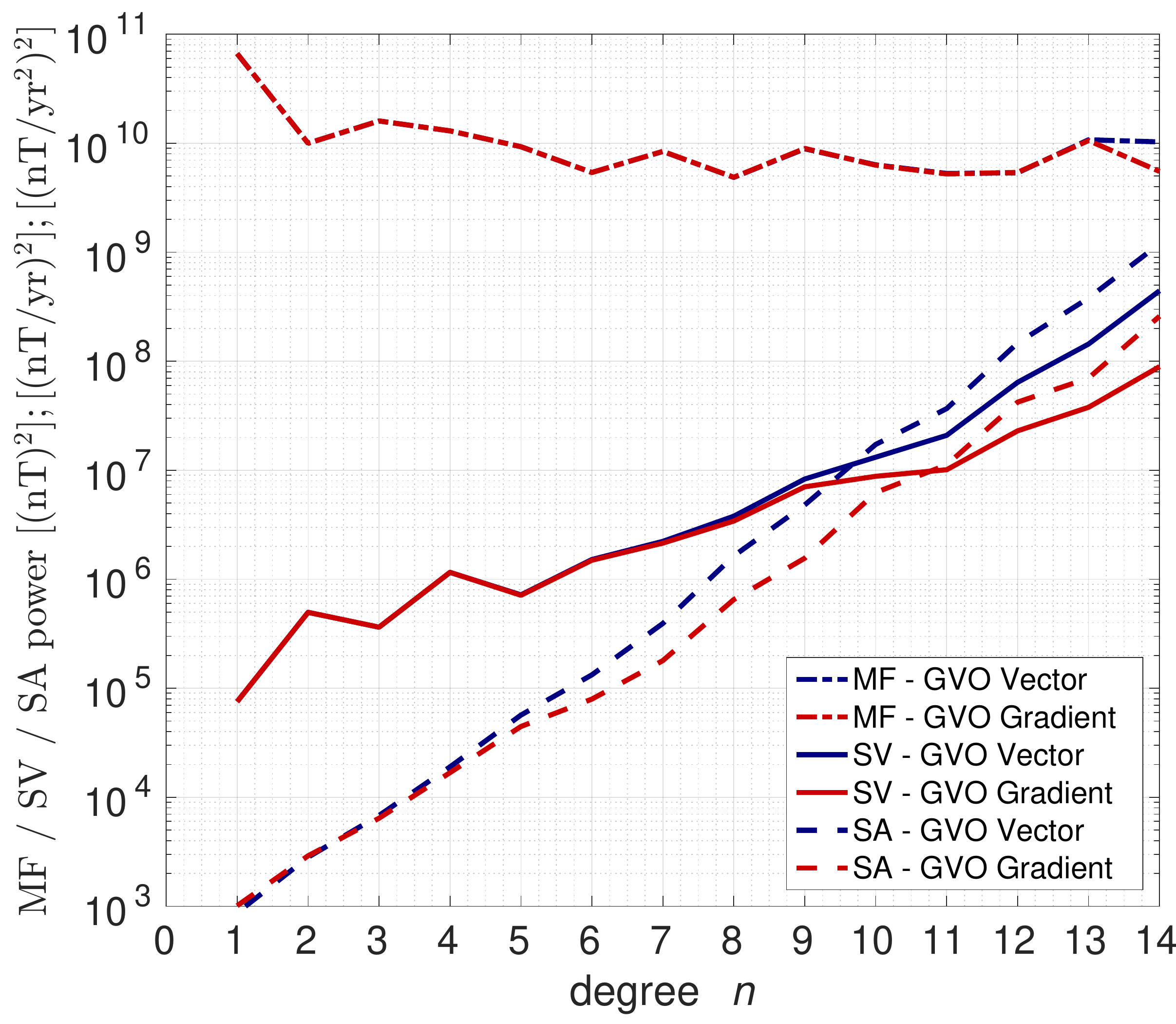} \includegraphics[width=0.49\textwidth]{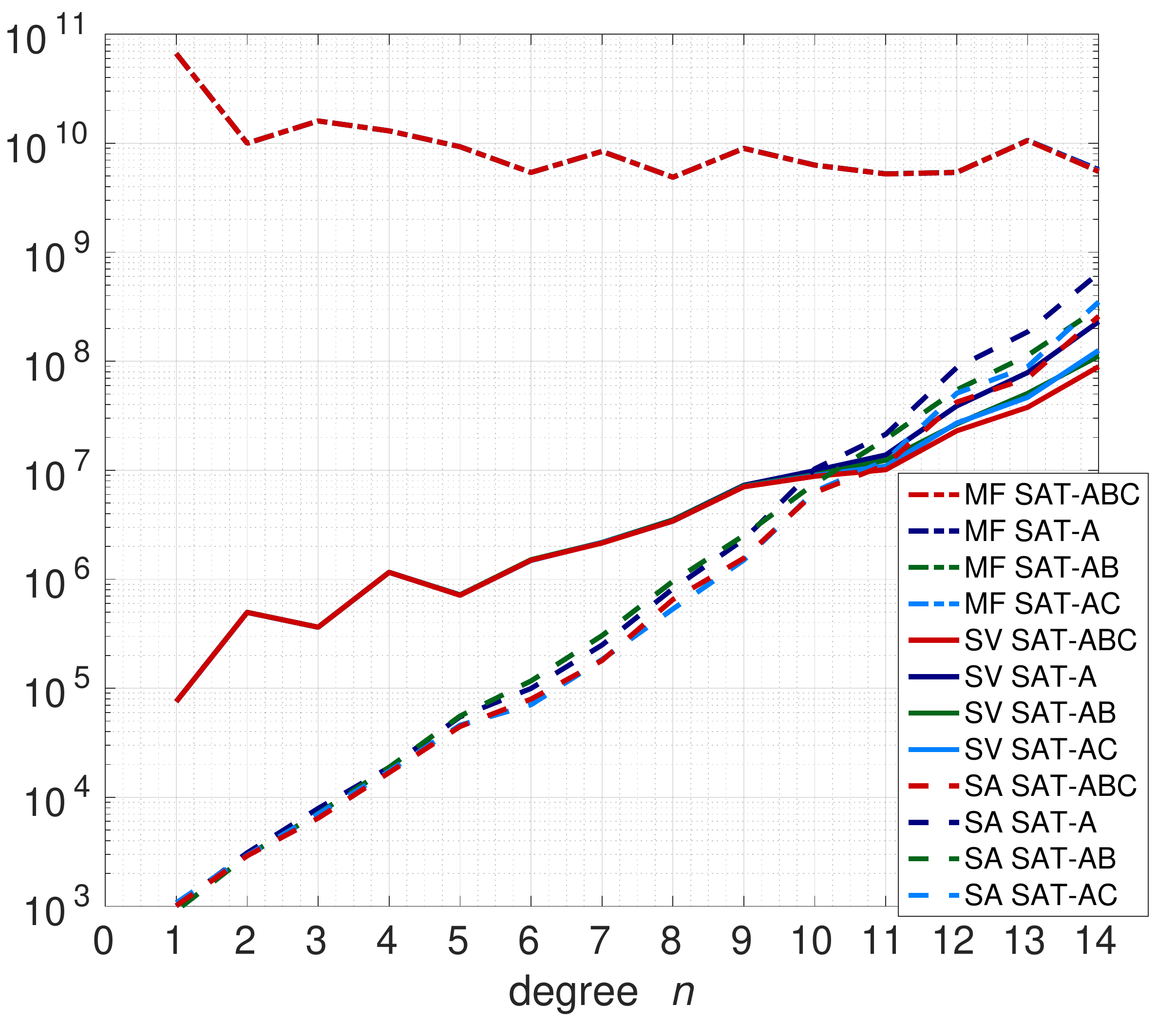}}
\caption{MF (\added{dashed-}dotted\deleted{ curves}), SV (\replaced{solid}{curves}) and SA (\replaced{dashed}{punctuated curves}) CMB mean power spectra of epoch models truncated at SH degree $n=14$. Left plot show spectra\deleted{l lines} from vector field (blue) and field gradient (red) GVOs, derived using all {\it Swarm} satellites. Right plot show \replaced{curves}{lines} based on gradient GVOs derived using: all {\it Swarm} satellites (red, SAT-ABC), {\it Swarm}-Alpha (dark blue, SAT-A), {\it Swarm}-Alpha and Bravo (green, SAT-AB) and {\it Swarm}-Alpha and Bravo (light blue, SAT-AC).}
\label{Fig:12}	
\end{figure} 
Next, we seek to further inspect and compare the SH models obtained from GVO \deleted{the field }gradient series with similar models obtained using more traditional GVO vector \replaced{component}{field} series. \added{The left plot of }Figure \ref{Fig:12} presents the mean of the MF (\added{dashed-}dotted\deleted{ curves}), SV (solid\deleted{ curves}) and SA (\replaced{dashed}{punctuated}\deleted{ curves}) \added{spatial Lowes-Mauersberger} power spectra \added{\citep{Hulot_etal_2015},} at the \added{Core-mantle boundary} \added{(}CMB\added{)} obtained \replaced{by computing the time-averaged spectrum taken over all the epoch spectral lines}{from the epoch-by-epoch SH models derived without applying any spatial or temporal regularization}. Spectral curves in blue and red are derived from GVO gradients and GVO vector series, respectively. The SV and SA power spectra derived from GVO gradient tensor series are seen to diverge less rapidly as compared to models derived from GVO vector series, and the SV/SA intersection happens at a slightly higher degree (\replaced{11}{10} compared to 9). This behaviour is consistent with the analyzes of \cite{kotsiaros_Olsen_2014}, who found that gradient \replaced{data}{observations} better constrain SV to higher SH degrees than vector \replaced{data}{observations}. \deleted{We find (not shown) that we can robustly map the SA at the CMB up to degree 7 using the 4-monthly gradient tensor element data. Although the CMB maps exhibit more noise due to the downward continuation of the field, they display the same distinct SA changes in the Pacific region as those appearing in Figure \ref{Fig:13}, thus supporting an internal origin of the 2017 SA impulse.} 

\added{In order to characterize the influence of satellite configuration on the GVO gradient series, we consider case studies consisting of: 1) a single satellite, 2) two satellites at different altitudes which would provide a better local-time coverage as compared to the first case, 3) East-West capability from side-by-side flying satellites. For each of these cases, we derived GVO gradient series using: 1) {\it Swarm} Alpha only (SAT-A), 2) both {\it Swarm} Alpha and Bravo (SAT-AB) covering altitude ranges of approx. 430-515km and 490-530km during 2014-2020, respectively, 3) {\it Swarm} Alpha and Charlie (SAT-AC) taking advantage of their East-West longitudinal separation of approx. $1.4^{\circ}$ (corresponding to 155km\deleted{)} at the equator\added{)} \citep{Olsen_etal_2015}. The right plot of Figure \ref{Fig:12} presents the time-averaged MF (\added{dashed-}dotted\deleted{ curves}), SV (solid\deleted{ curves}) and SA (\replaced{dashed}{punctuated curves}) spatial power spectra, derived from each of these three case. Note here, that all of the MF spectra overlap and are thus hidden by the red SAT-ABC spectrum. In addition, note that this red SAT-ABC spectrum in the right panel refers to the same model as the red MF-GVO Gradient spectrum in the left panel of Figure \ref{Fig:12}. We discuss these results and their implications in more detail in the next section.}

\begin{figure}
\centerline{\includegraphics[width=1.0\textwidth]{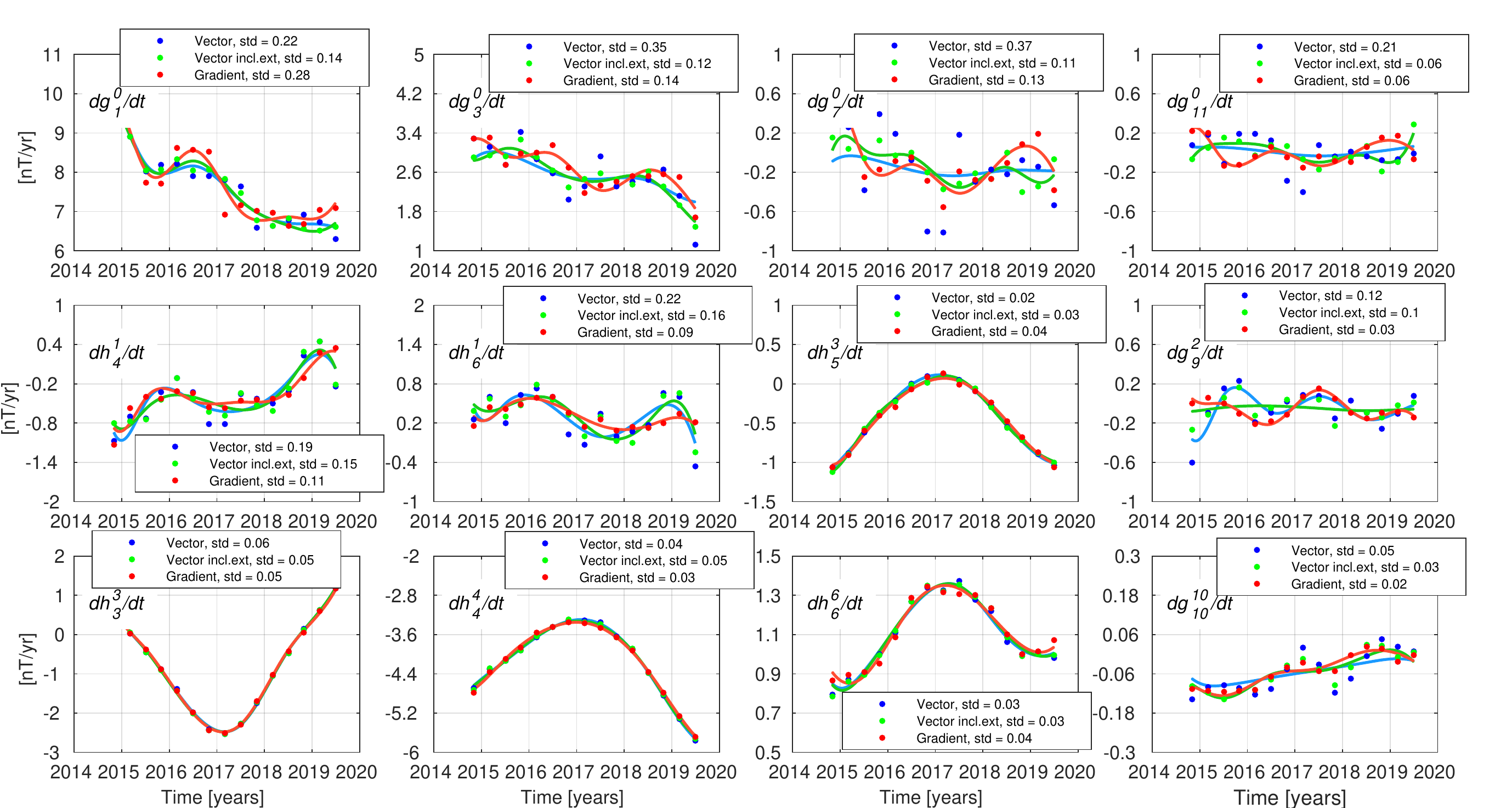}}
\caption{Series of first time derivatives for \replaced{some}{example} internal coefficients $dg_n^m/dt$ and $dh_n^m/dt$ derived from GVO vector (blue dots), GVO vector including an external SH expansion (green dots) and GVO gradient (red dots) data. Standard deviations of differences between the series and a GCV smoothing spline fit (solid curves) to the coefficients are given. Units are nT/yr.}
\label{Fig:14}	
\end{figure}

Investigating further these SH models, Figure \ref{Fig:14} shows the first time derivative of the internal expansion coefficients, computed based on simple first differences, derived from \added{internal models using} the GVO vector (blue\added{, and also including an external expansion, in green}) and GVO gradient (red) series \added{derived using all three {\it Swarm satellites}}. Example coefficients are shown for zonal, $m=0$, terms (top row), tesseral, $m \neq n$, terms (middle row) and sectorial, $m=n$, terms (bottom row). To quantify the scatter level in the epoch coefficient series, standard deviations between the coefficient series and GCV smoothing spline fits (solid curves) are given in each case. Although robust estimation has been \replaced{applied}{employed} when deriving these models, outliers can be seen in the both series. A change in the sign of the trend in the SV signal is evident, especially in the sectorial coefficients $\dot{h}_3^3,\dot{h}_4^4,\replaced{\dot{h}_6^6}{\dot{h}_5^5}$, but also in the $\dot{h}_5^3$ coefficient centered on 2017. The scatter level in the near zonal coefficients of the GVO vector series are reduced if an external SH expansion (in green dots and curve) is included as well. This points to the presence of external contamination in the vector series, which seems to be reduced in the gradient series.  

\begin{figure}
\centerline{\includegraphics[width=1.0\textwidth]{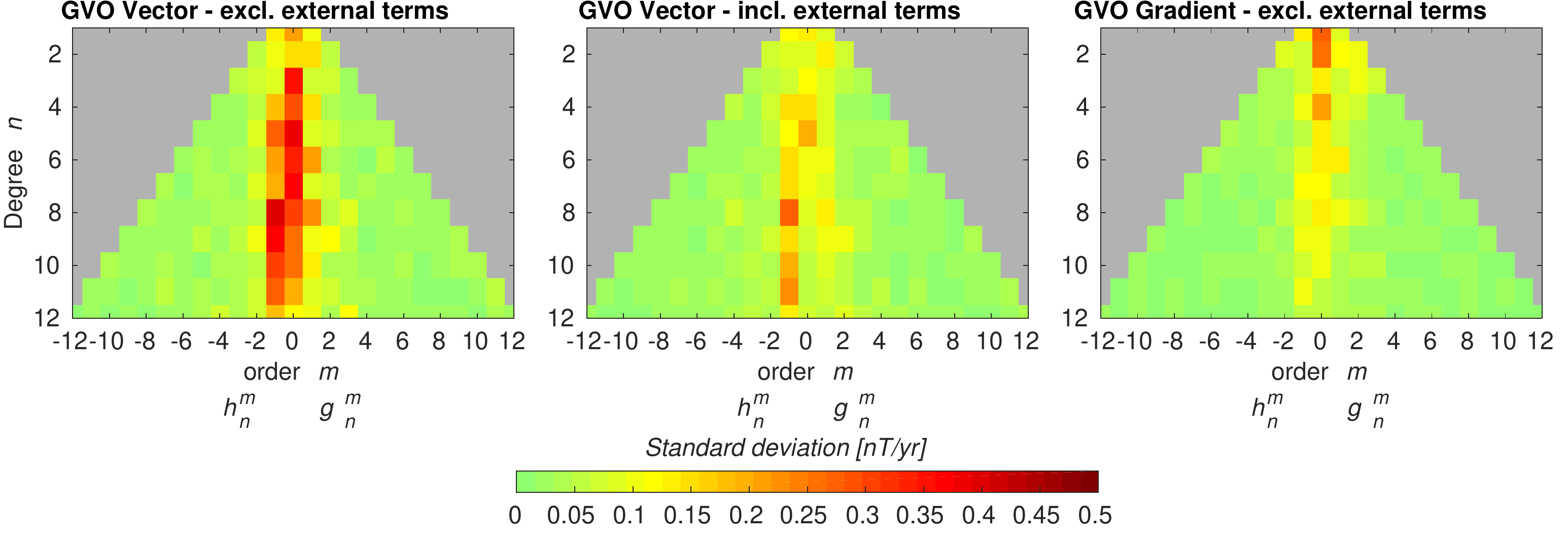}}
\caption{Standard deviations of differences between the first time derivative of internal SH model coefficient series and spline-fitted curves for each series derived from GVO vector data (left plot) and GVO gradient data (right plot), and GVO vector data including an external SH expansion (middle plot). Positive orders $m$ refer to the coefficients $dg_n^m/dt$, while negative orders refer to $dh_n^m/dt$ coefficients. Units are nT/yr.}	
\label{Fig:15}
\end{figure}

Figure \ref{Fig:15} collects such standard deviations for SH degrees and order up to 12, from models derived using the GVO vector (left plot) and GVO gradient (right plot) series. We generally find less scatter in the GVO gradient series, and especially in the zonal and near-zonal (where $m$ is close to zero) coefficients. For the sectorial terms the scatter levels are low for both series. Use of the vector series results in higher scatter levels for the near-zonal terms for degrees $n>2$ and orders $m \leq 2$. When also including an external SH expansion\added{, i.e. using both internal and external expansion terms up to degree 14,} for the GVO vector series (middle plot), we are able to reduce the scatter level in the near-zonal coefficients (middle plot of Figure \ref{Fig:15}), illustrating that using the GVO gradient elements in SH modelling helps in excluding external field signals. Note here that we focus on comparing SH models derived from GVO vector data with those derived solely using GVO gradient data (including an external SH expansion for the GVO gradient series would require vector information as well to obtaining a robust estimation). Global maps (not shown) show that much of the enhanced scatter is related to signals in polar regions being spuriously mapped into the internal field.

\begin{figure}
\centerline{\includegraphics[width=1.0\textwidth]{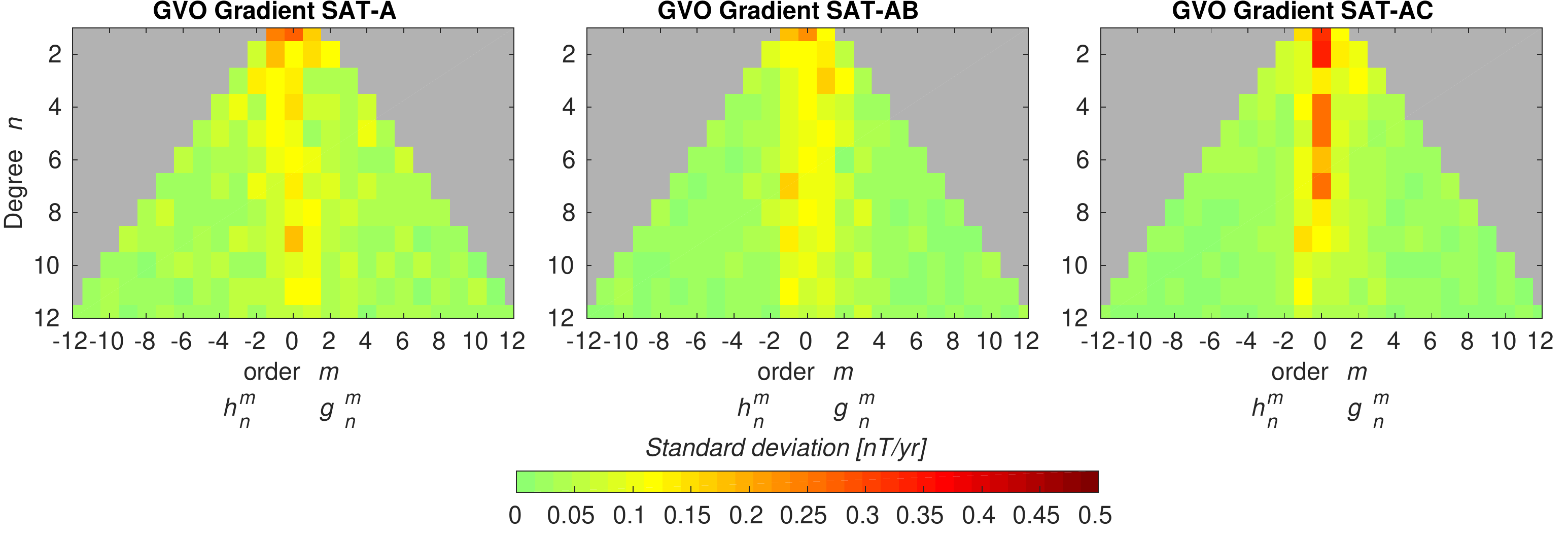}}
\caption{Standard deviations of differences between the first time derivative of internal SH model coefficient series and spline-fitted curves for each series based on GVO gradient series which has been derived using: {\it Swarm}-Alpha data (left plot, SAT-A), {\it Swarm} Alpha and Bravo data (center plot SAT-AB), and {\it Swarm} Alpha and Charlie data (right, SAT-AC). Positive orders $m$ refer to the coefficients $dg_n^m/dt$, while negative orders refer to $dh_n^m/dt$ coefficients. Units are nT/yr.}	
\label{Fig:16}
\end{figure}

\added{Finally, we show in Figure \ref{Fig:16}, the standard deviations from SH models based on the case study GVO gradient series, which were derived using {\it Swarm} Alpha only (case 1, left plot), {\it Swarm} Alpha and Bravo (case 2, center plot), and {\it Swarm} Alpha and Charlie (case 3, right plot).} 

\section{Discussion and Conclusions}
\label{sec:6}
In this study we have extended the existing GVO concept and derived time series of the second-order gradient tensor elements of the geomagnetic field at a global network of 300 locations. We have computed such GVO gradient time series from the mean and differences of vector magnetic field measurements, along track and in the east-west direction, from the low Earth orbiting CHAMP and {\it Swarm} satellites.

Inspecting SV gradient tensor elements for a GVO located above the Honolulu ground observatory we found evidence in the gradient series for a regional jerk-type event centered on 2017, observed as a characteristic "$V$" shaped change in the $d\left[\nabla B\right]_{rr}/dt$ and $d\left[\nabla B\right]_{r \theta}/dt$ elements, and as a "$\Lambda$" shape in the $d\left[\nabla B\right]_{\theta \theta}/dt$ and $d\left[\nabla B\right]_{\phi \phi}/dt$ elements. In the global GVO SV gradient element records, spanning the years from 2014 to 2020, we find evidence \replaced{of}{for} robust time variations in many of the tensor elements. In particular, intense fluctuations in the Pacific region confined in longitude, suggest a regionally localized geomagnetic impulse event taking place around 2017. This is consistent with ground observatory measurements of the SV of the radial magnetic field component at the Honolulu observatory  \cite[e.g.][]{Finlay_etal_2020,Sabaka_etal_2018}. By changing the geomagnetic quiet-time and local time selection criteria, we see little change in the amplitude of this jerk signal, supporting the hypothesis that the 2017 event is of internal origin. At the Earth's surface nearby patches of intense change in the SA gradient field, with opposite signs, occurs between 2015.5 and 2018.5. These are found to be limited to latitudes between $25^{\circ}$S to $25^{\circ}$N and to longitudes between $140^{\circ}$ to $220^{\circ}$E. In particular, two strong patches of change in the radial gradient of the radial field, with opposite signs, locate the centre of the 2017 jerk event to approximately $0^{\circ}$N and $170^{\circ}$E in the central Pacific. \added{We also find (not shown) that we can robustly map the change in SA at the CMB up to degree 7 using the 4-monthly gradient tensor element data. Although maps at the CMB exhibit enhanced noise due to downward continuation of the field they do display coherency regarding the distinctive SA changes observed in the Pacific region which are clearly related to those found at the Earth’s surface in Figure \ref{Fig:13}, further supporting an internal origin of the 2017 SA impulse.}

Various geophysical explanations of geomagnetic jerk events, similar to those we have highlighted here in the Pacific region, have been proposed. The possibilities still under discussion include equatorially trapped MAC waves in a possible stratified layer close to the core surface \citep{Buffett_Matsui_2019,ChiDuran_etal_2020} and equatorial focusing of hydrodynamic waves originating from turbulent convection deep within the core \citep{Aubert_Finlay_2019,Gerick_etal_2020}. In that connection the new concept of GVO gradient tensor time series may aid future studies of the appearance and dynamics of geomagnetic jerks, related changes in core flows and core dynamics via e.g. data assimilation. \deleted{We have shown that some GVO gradient tensor elements are less affected by correlated errors due to external field unmodelled signals, compared with vector field components. }In a follow-up study with \deleted{Prof. }K. Whaler (in prep), we \deleted{shall }present computations and investigations of core surface flows derived from GVO gradient tensor elements series, paying particular attention to the jerk in the Pacific region in 2017. 

\added{In addition to the signature of geomagnetic jerks, global maps of the GVO SV tensor element series, also show some example rapid smaller amplitude SV fluctuations within a few years, especially noticeable in the $d\left[\nabla B\right]_{rr}/dt$ element (see left panel of Figure \ref{Fig:10b}). Some of these small amplitude variations in the SV gradient series might be indicative of external field leakage. The most obvious sources of external field leakage are 1) rapid variations in polar latitude GVO series that persist to lower latitudes along the same meridional, which could be an indication of contamination by polar electrojet or field aligned currents, 2) rapid variations at mid or low latitude GVO series seen at all longitudes that could be caused by a signal having magnetospheric origin. We find no signs of distinct temporal variations along all longitudes at mid/low latitudes, suggesting that remaining magnetospheric disturbances are likely small. We do find some examples of rapid variations that seem to diminish towards lower latitudes, which could signify contamination from polar electrojet or field aligned currents systems. Longer time series are needed in order to study the origin of such variability in the GVO gradient series.}

In order to test for possible improvements in retrieving the SV signal using GVO gradient tensor data \replaced{as compared to GVO vector data}{alone}, we produced simple unregularized SH field models built from the GVO gradient and vector data derived using {\it Swarm} measurements. Comparing the power spectra of these models \added{shown in Figure \ref{Fig:12}} supports the findings of \cite{kotsiaros_Olsen_2014}, that harmonics of the SV above degree 6 can be better resolved when using gradient tensor data than using vector data. In particular, analysis of the first time derivatives of the SH coefficients as \replaced{presented}{shown} in Figure \ref{Fig:15}, shows that especially zonal and near-zonal harmonics of models derived from GVO gradients have less scatter compared to similar models derived from GVO vector data. 

\added{We find that the scatter level in the SV gradient tensor elements is higher during CHAMP times compared with {\it Swarm} times. The orbital configuration of the three {\it Swarm} satellites is advantageous for computing the gradient tensor as more data are available and {\it Swarm} Bravo, having a slightly higher altitude than Alpha and Charlie, provides information on the radial gradient which enables better potential determination. The right panel of Figure \ref{Fig:12} shows that having two satellites, one of which is at a higher orbit, slightly lowers the spatial power (green) as compared with having one satellite (dark blue). In addition, from Figure \ref{Fig:16}, having two satellites lowers the scatter of the tesseral coefficients for $m>2$. Considering instead two satellites in a side-by-side orbit, the associated spectral line (light blue in the right panel of Figure \ref{Fig:12}) is only marginally lower, but differences are most obvious from Figure \ref{Fig:16}, where the scatter level of the tesseral coefficients is smaller for orders $m>2$. The zonal terms for $n<7$ are found to have higher scatter, but this is expected as East-West information alone does not allow determination of zonal terms \citep{kotsiaros_Olsen_2012}. Comparing the three case studies with the original case of using all three {\it Swarm} satellites, the associated spatial power (red in the right panel of Figure \ref{Fig:12}) is lower, while the scatter level across all SH coefficients as shown in Figure \ref{Fig:15} is the lowest, except for lowest zonal terms which seems to be related to including the East-West gradient information as shown in the right plot of Figure \ref{Fig:16}. In order to mitigate such behavior of the low degree zonal terms, in future studies it might be worth exploring a Selective Infinite-Variance Weighting (SIVW) approach \citep{kotsiaros_Olsen_2014,Sabaka_etal_2013}, wherein weights are applied to data subsets that are more sensitive to certain parameter subsets. By this approach, the East-West data could in the future be given less weight with regard to parts of the GVO potential that provide information of the zonal terms. The larger misfits during CHAMP times, may also result from a closer proximity of the lower flying CHAMP satellite to ionospheric current systems. From our tests we therefore conclude that the {\it Swarm} satellite trio is advantageous for deriving GVO gradient series. Having satellites at different altitudes better fills the 3D space of the GVO data cylinder, improving the recovery of the gradient quantities, for example improving the determination of the tesseral harmonics. Using all three satellites, enhances the recovery of all harmonic coefficients, and is clearly superior to having a single satellite.}\deleted{The scatter levels for the SV of the gradient tensor elements are higher during CHAMP times than during {\it Swarm} times. The orbital configuration of the three {\it Swarm} satellites is clearly advantageous when computing the gradient tensor as more data are available and {\it Swarm} Bravo, having a slightly higher altitude than Alpha and Charlie, provides more information on the radial gradient and enables better potential determination and superior rms misfit statistics. Higher levels of scatter at polar latitudes are likely due to contaminating fields from polar current systems, while the generally larger misfits during CHAMP times at all latitudes as compared to {\it Swarm} times, are likely due to less complete data coverage and the closer proximity of the lower flying CHAMP satellite to ionospheric current systems. The SV gradient tensor elements $\left[ \nabla B\right]_{\theta \theta}$ and $\left[ \nabla B\right]_{r \theta}$, show lower levels of scatter compared to the other tensor elements, at both polar and at non-polar regions, correctly weighting the various components will be important for future applications.}

\section*{Availability of datasets and material}
The GVO gradient tensor data underlying this article and their associated uncertainty estimates are available from \url{https://data.dtu.dk/}, at \citep{Hammer_data}. The datasets used in this article are available in the following repositories: Swarm data are available from \url{https://earth.esa.int/web/guest/swarm/data-access}; CHAMP data are available from \url{https://isdc.gfz-potsdam.de/champ-isdc}; Ground observatory data are available from \url{ftp://ftp.nerc-murchison.ac.uk/geomag/Swarm/AUX_OBS/hour/}; The RC-index is available from \url{http://www.spacecenter.dk/files/magnetic-models/RC/}; The CHAOS-7\added{.2} model and its updates are available at \url{http://www.spacecenter.dk/files/magnetic-models/CHAOS-7/}; solar wind speed, interplanetary magnetic field, and Kp-index are available from \url{https ://omniw eb.gsfc.nasa.gov/ow.html}.
\begin{acknowledgments}
We wish to thank Gauthier Hulot and an anonymous reviewer for constructive comments that helped us improve the manuscript. We thank the GFZ German Research Centre for Geoscience for providing access to the CHAMP MAG-L3 data and the European Space Agency (ESA) for providing access to the {\it Swarm} L1b data. The high resolution 1-min OMNI data were provided by the Space Physics Data Facility (SPDF), NASA Goddard Space Flight Centre. We thank the staff of the geomagnetic observatories and the INTERMAGNET for providing high-quality observatory data. This project was partly supported by the European Research Council (ERC) under the European Union’s Horizon 2020 research and innovation programme (grant agreement No. 772561). In addition, this study was also partly funded by ESA through the Swarm DISC activities, contract no. 4000109587. 
\end{acknowledgments}
\bibliographystyle{gji}
\bibliography{VOgradients}

\begin{thebibliography}{49}
\expandafter\ifx\csname natexlab\endcsname\relax\def\natexlab#1{#1}\fi

\bibitem[Aubert \& Finlay(2019)]{Aubert_Finlay_2019}
Aubert, J. \& Finlay, C.~C., 2019.
\newblock Geomagnetic jerks and rapid hydromagnetic waves focusing at {E}arth's
  core surface, {\it Nature Geoscience\/}, {\bf 12}(5), 393--398.

\bibitem[Backus et~al.(1996)Backus, Parker, \& Constable]{Backus_etal_1996}
Backus, G., Parker, R., \& Constable, C., 1996.
\newblock {\it Foundations of Geomagnetism\/}, Cambridge Univ. Press, New York.

\bibitem[Barrois et~al.(2018)Barrois, Hammer, Finlay, Martin, \&
  Gillet]{Barrois_etal_2018}
Barrois, O., Hammer, M.~D., Finlay, C.~C., Martin, Y., \& Gillet, N., 2018.
\newblock Assimilation of ground and satellite magnetic measurements: inference
  of core surface magnetic and velocity field changes, {\it Geophys. J.
  Int.\/}, {\bf 215}, 695--712.

\bibitem[Beggan et~al.(2009)Beggan, Whaler, \& Macmillan]{Beggan_etal_2009}
Beggan, C.~D., Whaler, K.~A., \& Macmillan, S., 2009.
\newblock Biased residuals of core flow models from satellite-derived virtual
  observatories, {\it Geophys. J. Int.\/}, {\bf 177}(2), 463--475.

\bibitem[Bendat \& Piersol(2010)]{Bendat_Piersol_2010}
Bendat, J. \& Piersol, A., 2010.
\newblock {\it Random Data, Analysis and Measurement Procedures\/}, Wiley, New
  Jersey.

\bibitem[Buffett \& Matsui(2019)]{Buffett_Matsui_2019}
Buffett, B. \& Matsui, H., 2019.
\newblock Equatorially trapped waves in {E}arth's core, {\it Geophys. J.
  Int.\/}, {\bf 218}(2), 1210--1225.

\bibitem[Casotto \& Fantino(2009)]{Casotto_Fantino_2009}
Casotto, S. \& Fantino, E., 2009.
\newblock Gravitational gradients by tensor analysis with application to
  spherical coordinates, {\it Journal of Geodesy\/}, {\bf 83}(7), 621--634.

\bibitem[Chi-Dur{\'a}n et~al.(2020)Chi-Dur{\'a}n, Avery, Knezek, \&
  Buffett]{ChiDuran_etal_2020}
Chi-Dur{\'a}n, R., Avery, M.~S., Knezek, N., \& Buffett, B.~A., 2020.
\newblock Decomposition of geomagnetic secular acceleration into traveling
  waves using complex empirical orthogonal functions, {\it Geophys. Res.
  Lett.\/}, p. e2020GL087940.

\bibitem[Constable(1988)]{Constable_1988}
Constable, C.~G., 1988.
\newblock Parameter estimation in non-{G}aussian noise, {\it Geophys. J.
  Int.\/}, {\bf 94}(1), 131--142.

\bibitem[Domingos et~al.(2019)Domingos, Pais, Jault, \&
  Mandea]{Domingos_etal_2019}
Domingos, J., Pais, M.~A., Jault, D., \& Mandea, M., 2019.
\newblock Temporal resolution of internal magnetic field modes from satellite
  data, {\it Earth, Planets and Space\/}, {\bf 71}(1), 1--17.

\bibitem[Finlay(2019)]{Finlay_2019}
Finlay, C.~C., 2019.
\newblock Models of the main geomagnetic field based on multi-satellite
  magnetic data, {\it Ionospheric Multi-Spacecraft Analysis Tools: Approaches
  for Deriving Ionospheric Parameters\/}, {\bf 17}, 255.

\bibitem[Finlay et~al.(2017)Finlay, Lesur, Th{\'e}bault, Vervelidou,
  Morschhauser, \& Shore]{Finlay_etal_2017}
Finlay, C.~C., Lesur, V., Th{\'e}bault, E., Vervelidou, F., Morschhauser, A.,
  \& Shore, R., 2017.
\newblock Challenges handling magnetospheric and ionospheric signals in
  internal geomagnetic field modelling, {\it Space Science Reviews\/}, {\bf
  206}(1-4), 157--189.

\bibitem[Finlay et~al.(2020)Finlay, Kloss, Olsen, Hammer, T{\o}ffner-Clausen,
  Grayver, \& Kuvshinov]{Finlay_etal_2020}
Finlay, C.~C., Kloss, C., Olsen, N., Hammer, M.~D., T{\o}ffner-Clausen, L.,
  Grayver, A., \& Kuvshinov, A., 2020.
\newblock The {CHAOS}-7 geomagnetic field model and observed changes in the
  {S}outh {A}tlantic {A}nomaly, {\it Earth, Planets and Space\/}, {\bf 72}(1),
  1--31.

\bibitem[Gerick et~al.(2021)Gerick, Jault, \& Noir]{Gerick_etal_2020}
Gerick, F., Jault, D., \& Noir, J., 2021.
\newblock {F}ast {Q}uasi-{G}eostrophic {M}agneto-{C}oriolis {M}odes in the
  {E}arth's core, {\it Geophys. Res. Lett.\/}, {\bf 48}(4), e2020GL090803,
  doi:10.1029/2020GL090803.

\bibitem[Green \& Silverman(1993)]{Green_Silverman_1993}
Green, P.~J. \& Silverman, B.~W., 1993.
\newblock {\it Nonparametric regression and generalized linear models: a
  roughness penalty approach\/}, Chapman and Hall.

\bibitem[Hammer(2018)]{Hammer_2018}
Hammer, M.~D., 2018.
\newblock {\it Local Estimation of the {E}arth's Core Magnetic Field\/}, Ph.D.
  thesis, Technical University of Denmark.

\bibitem[Hammer \& Finlay(2019)]{Hammer_Finlay_2019}
Hammer, M.~D. \& Finlay, C.~C., 2019.
\newblock Local averages of the core--mantle boundary magnetic field from
  satellite observations, {\it Geophys. J. Int.\/}, {\bf 216}(3), 1901--1918.

\bibitem[Hammer et~al.(2021{\natexlab{a}})Hammer, Cox, Brown, Beggan, \&
  Finlay]{Hammer_etal_2021a}
Hammer, M.~D., Cox, G., Brown, W., Beggan, C.~D., \& Finlay, C.~C.,
  2021{\natexlab{a}}.
\newblock Geomagnetic {V}irtual {O}bservatories: monitoring geomagnetic secular
  variation with the {S}warm satellites, {\it Earth, Planets and Space\/}, {\bf
  73}(1), 1--22.

\bibitem[Hammer et~al.(2021{\natexlab{b}})Hammer, Finlay, \&
  Olsen]{Hammer_data}
Hammer, M.~D., Finlay, C.~C., \& Olsen, N., 2021{\natexlab{b}}.
\newblock Secular variation signals in magnetic field gradient tensor elements
  derived from satellite-based geomagnetic virtual observatories., Technical
  University of Denmark. Dataset. doi:10.11583/DTU.14695590.

\bibitem[Hammer et~al.(2021{\natexlab{c}})Hammer, Finlay, \&
  Olsen]{Hammer_etal_2021b}
Hammer, M.~D., Finlay, C.~C., \& Olsen, N., 2021{\natexlab{c}}.
\newblock Applications of {C}ryo{S}at-2 satellite magnetic data in studies of
  the {E}arth's core field variations, {\it Earth, Planets and Space\/}, {\bf
  73}(1), 1--22.

\bibitem[Huder et~al.(2020)Huder, Gillet, Finlay, Hammer, \&
  Tchoungui]{Huder_etal_2020}
Huder, L., Gillet, N., Finlay, C.~C., Hammer, M.~D., \& Tchoungui, H., 2020.
\newblock {COV}-{OBS}.x2: 180 years of geomagnetic field evolution from
  ground-based and satellite observations, {\it Earth, Planets and Space\/},
  {\bf 72}(1), 1--18.

\bibitem[Hulot et~al.(2015)Hulot, Sabaka, Olsen, \& Fournier]{Hulot_etal_2015}
Hulot, G., Sabaka, T.~J., Olsen, N., \& Fournier, A., 2015.
\newblock The present field, in {\em Treatise on Geophysics\/}, vol. 5, chapter
  5.02, ed. Kono, M., Elsevier Ltd.

\bibitem[Kloss \& Finlay(2019)]{Kloss_Finlay_2019}
Kloss, C. \& Finlay, C.~C., 2019.
\newblock Time-dependent low latitude core flow and geomagnetic field
  acceleration pulses, {\it Geophys. J. Int.\/}, {\bf 217}, 140--168.

\bibitem[Koop(1993)]{Koop_1993}
Koop, R., 1993.
\newblock {\it Global gravity field modelling using satellite gravity
  gradiometry\/}, Netherlands Geodetic Comission, Publications on Geodesy, New
  Series, Number 38, Delft, The Netherlands.

\bibitem[Kotsiaros(2012)]{Kotsiaros_2012}
Kotsiaros, S., 2012.
\newblock {\it Determination of {E}arth's magnetic field from satellite
  constellation magnetic field observations\/}, Ph.D. thesis, {DTU} {S}pace.

\bibitem[Kotsiaros(2016)]{Kotsiaros_2016}
Kotsiaros, S., 2016.
\newblock Toward more complete magnetic gradiometry with the {S}warm mission,
  {\it Earth, Planets and Space\/}, {\bf 68}(1), 1--13.

\bibitem[Kotsiaros \& Olsen(2012)]{kotsiaros_Olsen_2012}
Kotsiaros, S. \& Olsen, N., 2012.
\newblock The geomagnetic field gradient tensor, {\it GEM-International Journal
  on Geomathematics\/}, {\bf 3}(2), 297--314.

\bibitem[Kotsiaros \& Olsen(2014)]{kotsiaros_Olsen_2014}
Kotsiaros, S. \& Olsen, N., 2014.
\newblock End-to-end simulation study of a full magnetic gradiometry mission,
  {\it Geophys. J. Int.\/}, {\bf 196}(1), 100--110.

\bibitem[Kotsiaros et~al.(2015)Kotsiaros, Finlay, \&
  Olsen]{Kotsiaros_etal_2015}
Kotsiaros, S., Finlay, C., \& Olsen, N., 2015.
\newblock Use of along-track magnetic field differences in lithospheric field
  modelling, {\it Geophys. J. Int.\/}, {\bf 200}(2), 878--887.

\bibitem[Laundal \& Richmond(2017)]{Laundal_Richmond_2017}
Laundal, K.~M. \& Richmond, A.~D., 2017.
\newblock Magnetic coordinate systems, {\it Space Science Reviews\/}, {\bf
  206}(1-4), 27--59.

\bibitem[Leopardi(2006)]{Leopardi_2006}
Leopardi, P., 2006.
\newblock A partition of the unit sphere into regions of equal area and small
  diameter, {\it Electronic Transactions on Numerical Analysis\/}, {\bf
  25}(12), 309--327.

\bibitem[Mandea \& Olsen(2006)]{Mandea_Olsen_2006}
Mandea, M. \& Olsen, N., 2006.
\newblock A new approach to directly determine the secular variation from
  magnetic satellite observations, {\it Geophys. Res. Lett.\/}, {\bf 33}(15).

\bibitem[Nogueira et~al.(2015)Nogueira, Scharnagl, Kotsiaros, \&
  Schilling]{Nogueira_etal_2015}
Nogueira, T., Scharnagl, J., Kotsiaros, S., \& Schilling, K., 2015.
\newblock Net{S}at-4{G} {A} four nano-satellite formation for global
  geomagnetic gradiometry, in {\em Proceedings of 10th IAA Symposium on Small
  Satellites for Earth Observation\/}.

\bibitem[Olsen \& Kotsiaros(2011)]{Olsen_Kotsiaros_2011}
Olsen, N. \& Kotsiaros, S., 2011.
\newblock Magnetic satellite missions and data, in {\em Geomagnetic
  Observations and Models\/}, pp. 27--44, Springer.

\bibitem[Olsen \& Mandea(2007)]{Olsen_Mandea_2007}
Olsen, N. \& Mandea, M., 2007.
\newblock Investigation of a secular variation impulse using satellite data:
  The 2003 geomagnetic jerk, {\it Earth Planet. Sci. Lett.\/}, {\bf 255}(1),
  94--105.

\bibitem[Olsen \& Stolle(2012)]{Olsen_Stolle_2012}
Olsen, N. \& Stolle, C., 2012.
\newblock Satellite geomagnetism, {\it Annu. Rev. Earth Planet. Sci.\/}, {\bf
  40}, 441--465.

\bibitem[Olsen et~al.(2006)Olsen, L{\"u}hr, Sabaka, Mandea, Rother,
  T{\o}ffner-Clausen, \& Choi]{Olsen_etal_2006}
Olsen, N., L{\"u}hr, H., Sabaka, T.~J., Mandea, M., Rother, M.,
  T{\o}ffner-Clausen, L., \& Choi, S., 2006.
\newblock {CHAOS}—a model of the {E}arth's magnetic field derived from
  {CHAMP}, {{\O}}rsted, and {SAC-C} magnetic satellite data, {\it Geophys. J.
  Int.\/}, {\bf 166}(1), 67--75.

\bibitem[Olsen et~al.(2014)Olsen, L{\"u}hr, Finlay, Sabaka, Michaelis, Rauberg,
  \& T{\o}ffner-Clausen]{Olsen_etal_2014}
Olsen, N., L{\"u}hr, H., Finlay, C.~C., Sabaka, T.~J., Michaelis, I., Rauberg,
  J., \& T{\o}ffner-Clausen, L., 2014.
\newblock The {CHAOS}-4 geomagnetic field model, {\it Geophys. J. Int.\/}, {\bf
  197}(2), 815--827.

\bibitem[Olsen et~al.(2015)Olsen, Hulot, Lesur, Finlay, Beggan, Chulliat,
  Sabaka, Floberghagen, Friis-Christensen, Haagmans, et~al.]{Olsen_etal_2015}
Olsen, N., Hulot, G., Lesur, V., Finlay, C.~C., Beggan, C., Chulliat, A.,
  Sabaka, T.~J., Floberghagen, R., Friis-Christensen, E., Haagmans, R., et~al.,
  2015.
\newblock The {S}warm {I}nitial {F}ield {M}odel for the 2014 geomagnetic field,
  {\it Geophys. Res. Lett.\/}, {\bf 42}(4), 1092--1098.

\bibitem[Olsen et~al.(2017)Olsen, Ravat, Finlay, \& Kother]{Olsen_etal_2017}
Olsen, N., Ravat, D., Finlay, C.~C., \& Kother, L.~K., 2017.
\newblock {LCS}-1: a high-resolution global model of the lithospheric magnetic
  field derived from {CHAMP} and {S}warm satellite observations, {\it Geophys.
  J. Int.\/}, {\bf 211}(3), 1461--1477.

\bibitem[Reed(1973)]{Reed_1973}
Reed, G., 1973.
\newblock {\it Application of kinematical geodesy for determining the shorts
  wavelength component of the gravity field by satellite gradiometry\/}, Ph.D.
  thesis, The Ohio State University, Dept. of Geod Science, Rep. No. 201,
  Columbus, Ohio.

\bibitem[Riley et~al.(2004)Riley, Hobson, \& Bence]{Riley_etal_2004}
Riley, K.~F., Hobson, M., \& Bence, S., 2004.
\newblock {\it Mathematical Methods for Physics and Engineering\/}, Cambridge
  University Press, Cambridge.

\bibitem[Ritter et~al.(2004)Ritter, L{\"u}hr, Maus, \&
  Viljanen]{Ritter_etal_2004}
Ritter, P., L{\"u}hr, H., Maus, S., \& Viljanen, A., 2004.
\newblock High-latitude ionospheric currents during very quiet times: their
  characteristics and predictability, {\it Annales Geophysicae\/}, {\bf 22},
  2001--2014.

\bibitem[Rogers et~al.(2019)Rogers, Beggan, \& Whaler]{Rogers_etal_2019}
Rogers, H.~F., Beggan, C.~D., \& Whaler, K.~A., 2019.
\newblock Investigation of regional variation in core flow models using
  spherical slepian functions, {\it Earth, Planets and Space\/}, {\bf 71}(1),
  19.

\bibitem[Sabaka et~al.(2010)Sabaka, Hulot, \& Olsen]{Sabaka_etal_2010}
Sabaka, T.~J., Hulot, G., \& Olsen, N., 2010.
\newblock Mathematical properties relevant to geomagnetic field modeling, in
  {\em Handbook of Geomathematics\/}, pp. 503--538, Springer.

\bibitem[Sabaka et~al.(2013)Sabaka, T{\o}ffner-Clausen, \&
  Olsen]{Sabaka_etal_2013}
Sabaka, T.~J., T{\o}ffner-Clausen, L., \& Olsen, N., 2013.
\newblock Use of the {C}omprehensive {I}nversion method for {S}warm satellite
  data analysis, {\it Earth, Planets and Space\/}, {\bf 65}(11), 1201--1222.

\bibitem[Sabaka et~al.(2018)Sabaka, T{\o}ffner-Clausen, Olsen, \&
  Finlay]{Sabaka_etal_2018}
Sabaka, T.~J., T{\o}ffner-Clausen, L., Olsen, N., \& Finlay, C.~C., 2018.
\newblock A {C}omprehensive {M}odel of the {E}arth's {M}agnetic {F}ield
  {D}etermined {F}rom 4 {Y}ears of {S}warm {S}atellite {O}bservations, {\it
  Earth, Planets and Space\/}, {\bf 70}(1), 1--26.

\bibitem[Shore(2013)]{Shore_2013}
Shore, R.~M., 2013.
\newblock {\it An improved description of {E}arth's external magnetic fields
  and their source regions using satellite data\/}, Ph.D. thesis, The
  University of Edinburgh.

\bibitem[Tscherning(1976)]{Tscherning_1976}
Tscherning, C., 1976.
\newblock Computation of the second-order derivatives of the normal potential
  based on the representation by a {L}egendre series, {\it Manuscripta
  geodaetica\/}, {\bf 1}, 71--92.

\end{thebibliography}
\clearpage
\appendix
\section{The Magnetic Field Gradient Tensor and General Coordinate Transformations}
\label{app:A}
Here we provide details on the magnetic gradient tensor and how its elements transform between different coordinate systems. In particular, we are interested in the transformation \deleted{relations }between \deleted{the tensor components of }the local topocentric Cartesian coordinate system described in Section \ref{sec:4.1} and the spherical \added{polar} coordinate system. Formulations from gravimetry of the gravitational gradient tensor (also referred to as the Marussi tensor) can be found in \cite{Reed_1973,Koop_1993,Casotto_Fantino_2009,Tscherning_1976}. Here we follow the notation of \cite{Casotto_Fantino_2009}, which is inspired by common usage in general relativity. The reader should however take care concerning the differences between the magnetic and gravity cases, and in particular, of the coordinate systems adopted, i.e. their orientation and whether they are left- or right-handed systems. 

Referring to a point $P$ (which would denote a given GVO target point), the usual geocentric system is given by the Cartesian coordinates as $\tilde{x}^p=(\tilde{x},\tilde{y},\tilde{z})$ and by the spherical polar coordinates as $(r,\theta,\phi)$, where $\theta$ is the colatitude. The geocentric system can be described by the Cartesian unit vectors $(\mathbf{\hat{i}_1},\mathbf{\hat{i}_2},\mathbf{\hat{i}_3})$ denoting the basis $i_p$. At $P$ a local Cartesian coordinate system $(z,x,y)$ is defined by the basis $\mathbf{e}_p$ where $p=1,2,3$, which is same one as used in GVO method, see Section \ref{sec:4.1}. This covariant right-handed orthogonal basis is determined by the components of the partial derivatives of the position vector $\mathbf{r}$ as: $\mathbf{e}_1=\partial \mathbf{r} / \partial r$ pointing radially outwards, $\mathbf{e}_2=\partial \mathbf{r} / \partial \theta$ pointing to the south and $\mathbf{e}_3=\partial \mathbf{r} / \partial \phi$ pointing to the east, i.e. similar to the spherical polar basis vector at the target point $P$. \added{The normalized basis vectors for the local system are given as $\hat{\mathbf{e}}_1=\mathbf{e}_1$, $\hat{\mathbf{e}}_2=(1/r)\mathbf{e}_2$, $\hat{\mathbf{e}}_3=1/(r \mathrm{sin}\, \theta) \mathbf{e}_1$, such that the position vector can be written as $\mathbf{r}=z\hat{\mathbf{e}_1}+x\hat{\mathbf{e}_2}+y\hat{\mathbf{e}_3}$}. Notice that while the basis vectors $\mathbf{i}_p$ are constant in magnitude and direction, the basis vectors $\hat{\mathbf{e}}_p$ have constant magnitude but their directions vary (the same goes for the spherical basis vectors). Thus when computing the spatial derivatives of a vector, the basis vectors also need\deleted{s} to be differentiated as \replaced{they}{these} depend on position. The position vector from origin $O$ to the point $P$ can be written \replaced{in}{by the} geocentric Cartesian coordinates as \citep{Riley_etal_2004}

\begin{equation}
\mathbf{r}  = \tilde{x}\mathbf{\hat{i}_1}+\tilde{y}\mathbf{\hat{i}_2}+\tilde{z}\mathbf{\hat{i}_3} = \tilde{x}^p i_p \; , \label{eq:position_vec}
\end{equation}

where the summation convention has been used. The Cartesian coordinates are related to the spherical coordinates via \citep[p.~363]{Riley_etal_2004}

\begin{equation}
\mathbf{r} = \begin{bmatrix} \tilde{x} \\ \tilde{y} \\ \tilde{z} \end{bmatrix} = r \begin{bmatrix} \mathrm{sin}\theta\mathrm{cos}\phi \\ \mathrm{sin}\theta\mathrm{sin}\phi \\ \mathrm{cos}\theta \end{bmatrix} \; .\label{eq:position_coor}
\end{equation}

The magnetic scalar potential, $V$, can be considered as a tensor of zero-order (or rank). The gradient operator in the generalized coordinates $u^p$, where $p=1,2,3$, having the covariant basis $\mathbf{e}_p=\partial \mathbf{r}/\partial u^p$ and contravariant basis $\mathbf{e}^p$ can be defined as \citep{Riley_etal_2004,Casotto_Fantino_2009}

\begin{equation}
\nabla = \mathbf{e}^p \frac{\partial}{\partial u^p} \; .  \label{eq:nabla}
\end{equation}

Applying the gradient operator to the potential generates a new tensor of one order higher, which is the first-order tensor (vector) describing the magnetic field

\begin{equation}
\replaced{\nabla V = V_{,p}\mathbf{e}^p}{\nabla V = \frac{1}{h_p} V_p \mathbf{e}^p} \; ,  \label{eq:nabla_v}
\end{equation}

where we use the \added{comma} notation $V_{\added{,}p} = \partial V / \partial u^p$ \added{to denote partial differentiation}. \added{The components that arise by applying an operator such as eq.\eqref{eq:nabla} which is represented in the covariant basis, are here referred to as the “natural” components. In contrast, describing the gradient in terms of the normalized basis vectors, the "physical" components of the first order gradient can be expressed as} 

\begin{equation}
V_{,p}^* = \frac{1}{h_p} \frac{\partial V}{\partial u^p } = \frac{1}{h_p} V_{,p} \; .  \label{eq:nabla_v_phys}
\end{equation}

\added{That is, we use the "*" notation to denote the physical components represented in the normalized basis, and distinguish it from the natural components.} The metric scale factor $h_p$ is determined by the elements of the metric tensor $g_{pq}=\mathbf{e}_p \cdot \mathbf{e}_q$ (which completely characterize any curvilinear coordinate system) as $h_p=\sqrt{g_{pp}}$. Note that the metric tensor also facilitates the conversion between covariant and contravariant bases \citep{Riley_etal_2004,Casotto_Fantino_2009}. \deleted{Here we follow \hbox{\cite{Casotto_Fantino_2009}} and denote the actual elements of the first-order gradient tensor by a semicolon notation} 
\deleted{
\begin{equation}
\deleted{V_{;p} = \frac{1}{h_p} V_p}
\end{equation}
}
\deleted{in order to distinguish them from first order derivatives.}
Applying the gradient operator to eq.\eqref{eq:nabla}, produces the second-order gradient operator 

\begin{align}
\nabla \nabla &= \mathbf{e}^q \frac{\partial}{\partial u^q} \left( \mathbf{e}^p \frac{\partial}{\partial u^p}  \right) \nonumber  \\
              &= \mathbf{e}^p \mathbf{e}^q \left(\frac{\partial^2}{\partial u^p \partial u^q} - \Gamma_{pq}^s \frac{\partial}{\partial u^s}  \right) , \label{eq:nabla_nabla}
\end{align}

where $\Gamma_{pq}^s$ denotes the Christoffel's symbols of the second kind (an array of numbers describing the derivatives of the covariant basis vector along that same basis), which can be expressed in terms of the metric tensor as \citep[p.~814]{Riley_etal_2004}

\begin{equation}
\Gamma_{pq}^s = \frac{1}{2} g^{st} \left( \frac{\partial g_{qt}}{\partial u^p} + \frac{\partial g_{pt}}{\partial u^q} - \frac{\partial g_{pq}}{\partial u^t} \right) \; . \label{eq:christoffel}
\end{equation}

Applying the operator eq.\eqref{eq:nabla_nabla} to the magnetic potential $V$ generates the second-order magnetic gradient tensor elements \deleted{(again adopting the semicolon notation in order to distinguish tensor elements e.g $V_{;rr}$ from the second derivative $V_{rr}=\partial^2 V / \partial r^2$),} which can be written using the Christoffel's symbols  

\begin{equation}
V_{;pq} = \deleted{\frac{1}{h_p h_q}} \left( \frac{\partial^2 V}{\partial u^p \partial u^q} - \Gamma_{pq}^s \frac{\partial V}{\partial u^s}  \right)  \; . \label{eq:nabla_nabla_V}
\end{equation}

\added{
Here the semicolon notation is used to denote covariant differentiation. In addition, indices which occurs after a semicolon, means differentiation. In physical components eq.\eqref{eq:nabla_nabla_V} is written as}

\begin{equation}
\added{V_{;pq}^* = \frac{1}{h_p h_q} V_{;pq} \; .} \label{eq:nabla_nabla_V_phys}
\end{equation}

\added{Recall here, that the "*" denotes the physical components.} An essential aspect of the first- and second-order tensors is how their \added{physical} elements \replaced{$V_{,p'}^*$}{$V_{;p'}$} or \replaced{$V^*_{;p'q'}$}{$V_{;p'q'}$} in one coordinate system $u^{p'}$ transforms to a new coordinate system $u^p$ \citep[p.~811]{Casotto_Fantino_2009,Riley_etal_2004}

\begin{align}
\replaced{V_{,p}^*}{V_{;p}}  &= \frac{h_{p'}}{h_p} \frac{\partial u^{p'}}{\partial u^p}\replaced{V_{,p'}^*}{V_{;p'}}  \label{eq:nabla_nabla_Trans1} \\
\replaced{V^*_{;pq}}{V_{;pq}}  &= \frac{h_{p'}}{h_p} \frac{h_{q'}}{h_q} \frac{\partial u^{p'}}{\partial u^p} \frac{\partial u^{p'}}{\partial u^q} \replaced{V^*_{;p'q'}}{V_{;p'q'}}  \; , \label{eq:nabla_nabla_Trans2}
\end{align}

where the partial derivatives $\partial u^{p'}/\partial u^p$ are expressed by the Jacobian matrix. The Jacobian matrix times the metric scale factor term, can be regarded as a rotation matrix such that we may re-write eqs.\eqref{eq:nabla_nabla_Trans1} and \eqref{eq:nabla_nabla_Trans2} 

\begin{align}
\replaced{V^*_{,p}}{V_{;p}}  &= \replaced{V^*_{,p'}}{V_{;p'}} R       \label{eq:nabla_nabla_Trans1_2} \\
\replaced{V^*_{;pq}}{V_{;pq}} &= R \replaced{V^*_{;p'q'}}{V_{;p'q'}} R^T  \label{eq:nabla_nabla_Trans2_2} \;,
\end{align}

having the transformation matrix determined as

\begin{equation}
R = \frac{\partial u^{p'}}{\partial u^p} D \; , \label{eq:R_matrix}
\end{equation}

where $D = diag(h_{p'}/h_p)=diag(h_{1'}/h_1,h_{2'}/h_2,h_{3'}/h_3)$ is a diagonal $3 \times 3$ matrix of the scale factor ratios between the two coordinate systems \added{using the physical components}. Thus equations \eqref{eq:nabla_nabla_Trans1} and \eqref{eq:nabla_nabla_Trans2} (equivalently eqs.\eqref{eq:nabla_nabla_Trans1_2} and \eqref{eq:nabla_nabla_Trans2_2}) allow us to transform the tensors in one coordinate system, for instance the global $(\tilde{x},\tilde{y},\tilde{z})$, to another, for instance $(r,\theta,\phi)$. \added{Note here, that for the equivalence of eqs.\eqref{eq:nabla_nabla_Trans1_2} and \eqref{eq:nabla_nabla_Trans2_2} in the natural components, the transformation matrix is defined by eq.\eqref{eq:R_matrix} but without the matrix $D$, i.e. without the metric correction.} Let us now consider the two transformations:
\begin{itemize}
\item[a)] Transformation from the global Cartesian $(\tilde{x},\tilde{y},\tilde{z})$ to the spherical system $(r,\theta,\phi)$
\item[b)] Transformation from the spherical system $(r,\theta,\phi)$ to the local system \replaced{$(z,x,y)$}{ $(x,y,z)$}
\end{itemize}
First, we specify the inner products of the basis vectors, the covariant metric tensors for the Cartesian system \added{, which is valid for both the global and local Cartesian coordinate systems having basis vectors $(\hat{\mathbf{i}}_1,\hat{\mathbf{i}}_2,\hat{\mathbf{i}}_3)$ and $(\hat{\mathbf{e}}_1,\hat{\mathbf{e}}_2,\hat{\mathbf{e}}_3)$, respectively}

\begin{equation}
g_{pq} = 
\left[
\begin{array}{ccc}
1  & 0 & 0 \\
0  & 1 & 0 \\ 
0  & 0 & 1  
\end{array}
\right] \; .\label{eq:metric_cartesian}
\end{equation}

and the spherical system, \added{having the un-normalized basis vectors $\mathbf{e}_p$, i.e. $(\mathbf{e}_r,\mathbf{e}_{\theta},\mathbf{e}_{\phi})$}

\begin{equation}
g_{pq} = 
\left[
\begin{array}{ccc}
1  & 0 & 0 \\
0  & r^2 & 0 \\ 
0  & 0 & r^2\mathrm{sin}^2\theta  
\end{array}
\right] \; . \label{eq:metric_spherical}
\end{equation}

\added{Note, that the un-normalized vectors $(\mathbf{e}_r,\mathbf{e}_{\theta},\mathbf{e}_{\phi})$ are the same as the vectors of the basis $\mathbf{e}_1$, $\mathbf{e}_2$ and $\mathbf{e}_3$ mentioned in the introduction of this appendix.} Thus the metric scale factors of the Cartesian system become 

\begin{equation}
h_{\tilde{x}} = 1, \quad h_{\tilde{y}} = 1, \quad h_{\tilde{z}} = 1 \; , \label{eq:scale_cartesian}
\end{equation}

and for the spherical system

\begin{equation}
h_r = 1, \quad h_{\theta} = r, \quad h_{\phi} = r \mathrm{sin}\theta  \; .\label{eq:scale_spherical}
\end{equation}

The Christoffel's symbols \added{for the spherical coordinate system} determined by eq.\eqref{eq:christoffel} yields 27 values of which 9 are non-zero

\begin{align}
\Gamma_{pq}^1 & = 
\left[
\begin{array}{ccc}
0  & 0  & 0 \\
0  & -r & 0 \\ 
0  & 0  & -r\mathrm{sin}^2\theta 
\end{array}
\right]   \nonumber \\
\Gamma_{pq}^2 & =
\left[
\begin{array}{ccc}
0            & \frac{1}{r}  & 0 \\
\frac{1}{r}  & 0            & 0 \\ 
0            & 0            & -\mathrm{cos}\theta\mathrm{sin}\theta 
\end{array}
\right]   \nonumber \\
\Gamma_{pq}^3 & =
\left[
\begin{array}{ccc}
0            & 0                                              & \frac{1}{r} \\
0            & 0                                              & \frac{\mathrm{cos}\theta}{\mathrm{sin}\theta} \\ 
\frac{1}{r}  & \frac{\mathrm{cos}\theta}{\mathrm{sin}\theta}  & 0 
\end{array}
\right]   \;  .   \label{eq:christoffel_spherical}
\end{align}

Note that for \added{both the global and local}\deleted{the} Cartesian system\added{s,} the Christoffel's symbols are zero as the metric tensor is the identity matrix. In case a) the Jacobian matrix between the spherical coordinates $u^p=(r,\theta,\phi)$ and the Cartesian coordinates $u^{p'}=\tilde{x}^p=(\tilde{x},\tilde{y},\tilde{z})$ is \added{derived considering the position vector in eq.\eqref{eq:position_coor}}

\begin{equation}
\left( \frac{\partial u^{p'}}{\partial u^p} \right) = \frac{\partial(\tilde{x},\tilde{y},\tilde{z})}{\partial(r,\theta,\phi)}  
                    =\begin{pmatrix}                                                   
\frac{\partial \tilde{x}}{\partial r}  & \frac{\partial \tilde{x}}{\partial \theta} & \frac{\partial \tilde{x}}{\partial \phi} \\
\frac{\partial \tilde{y}}{\partial r}  & \frac{\partial \tilde{y}}{\partial \theta} & \frac{\partial \tilde{y}}{\partial \phi} \\ 
\frac{\partial \tilde{z}}{\partial r}  & \frac{\partial \tilde{z}}{\partial \theta} & \frac{\partial \tilde{z}}{\partial \phi}                                                  
                    \end{pmatrix}  
                    =\begin{pmatrix}  
\mathrm{sin}\theta\mathrm{cos}\phi  & r\mathrm{cos}\theta\mathrm{cos}\phi & -r\mathrm{sin}\theta\mathrm{sin}\phi \\
\mathrm{sin}\theta\mathrm{sin}\phi  & r\mathrm{cos}\theta\mathrm{sin}\phi & r\mathrm{sin}\theta\mathrm{cos}\phi \\ 
\mathrm{cos}\theta                  & -r\mathrm{sin}\theta                & 0 
                   \end{pmatrix} \; .\label{eq:jacobian_globalTOspher}
\end{equation}

\replaced{In }{while in }case b) the Jacobian matrix between the local Cartesian coordinates $u^{p}=(z,x,y)$ and the spherical coordinates $u^{p'}=(r,\theta,\phi)$ is \added{found by considering the position vector in the local Cartesian basis}

\begin{equation}
\mathbf{r} = z \hat{\mathbf{e}}_1 + x \hat{\mathbf{e}}_2 + y \hat{\mathbf{e}}_3 = z \frac{\partial \mathbf{r}}{\partial r}+ \frac{x}{r} \frac{\partial \mathbf{r}}{\partial \theta} + \frac{y}{r\mathrm{sin}\,\theta} \frac{\partial \mathbf{r}}{\partial \phi} \; . \label{eq:pos_vec_2}
\end{equation}

\added{From this, we can compute the Jacobian matrix}

\begin{equation}
\left( \frac{\partial u^{p'}}{\partial u^p} \right) = \frac{\partial(r,\theta,\phi)}{\partial(z,x,y)}  
                    =\begin{pmatrix}                                                   
\frac{\partial r     }{\partial z}  & \frac{\partial r     }{\partial x} & \frac{\partial r     }{\partial y} \\
\frac{\partial \theta}{\partial z}  & \frac{\partial \theta}{\partial x} & \frac{\partial \theta}{\partial y} \\ 
\frac{\partial \phi  }{\partial z}  & \frac{\partial \phi  }{\partial x} & \frac{\partial \phi  }{\partial y}                                                  
                    \end{pmatrix}  
                    =\begin{pmatrix}  
1  & 0 & 0 \\
0  & \replaced{\frac{1}{r}}{1} & 0 \\ 
0  & 0 & \replaced{\frac{1}{r \mathrm{sin}\,\theta}}{1}
                   \end{pmatrix} \; .\label{eq:jacobian_spherTOlocal}
\end{equation}

\deleted{Considering case a), we use eqs.\eqref{eq:nabla_nabla_Trans1_2} and \eqref{eq:nabla_nabla_Trans2_2} to obtain the relations written here in matrix form}
\deleted{
\begin{align}
\deleted{\frac{\partial V}{\partial (r,\theta,\phi)}}     &\deleted{= \frac{\partial V}{\partial (\tilde{x},\tilde{y},\tilde{z})} R  }  \\
\deleted{\frac{\partial^2 V}{\partial (r,\theta,\phi)^2}} &\deleted{= R \frac{\partial V}{\partial (\tilde{x},\tilde{y},\tilde{z})^2} R^T \; ,  } 
\end{align}
}
\deleted{where $R$ is determined by eq.\eqref{eq:R_matrix} using eqs.\eqref{eq:scale_cartesian}, \eqref{eq:scale_spherical} and \eqref{eq:jacobian_globalTOspher}.}

\deleted{Likewise considering case b), we use the relations eqs.\eqref{eq:nabla_nabla_Trans1_2} and \eqref{eq:nabla_nabla_Trans2_2} written here in matrix form}
\deleted{
\begin{align}
\deleted{\frac{\partial V}{\partial (z,x,y)} }    &\deleted{= \frac{\partial V}{\partial (r,\theta,\phi)} R  } \\
\deleted{\frac{\partial^2 V}{\partial (z,x,y)^2}} &\deleted{= R \frac{\partial V}{\partial (r,\theta,\phi)^2} R^T \; , }
\end{align}
}
\deleted{where $R$ is determined by eq.\eqref{eq:R_matrix} using eqs.\eqref{eq:scale_cartesian}, \eqref{eq:scale_spherical} and \eqref{eq:jacobian_spherTOlocal}.}

\replaced{The}{Here the} first-order tensor (i.e. the magnetic field vector) in the spherical polar coordinates is given by eq.\eqref{eq:nabla_v_phys} using the metric scale factors eq.\eqref{eq:scale_spherical}

\begin{equation} 
\nabla V  = V_{\added{,}r} \mathbf{\hat{e}_r} +\frac{1}{r}V_{\added{,}\theta} \mathbf{\hat{e}_{\theta}} +\frac{1}{r\mathrm{sin}\theta}V_{\added{,}\phi} \mathbf{\hat{e}_{\phi}} \; .
\end{equation}

\added{The physical components of first-order magnetic tensor elements in the local Cartesian system\deleted{ by}, can be derived from eqs.\eqref{eq:nabla_nabla_Trans1_2} and \eqref{eq:R_matrix}, using the Jacobian matrix eq.\eqref{eq:jacobian_spherTOlocal}. Here the transformation matrix $R$ becomes the identity matrix because the Jacobian matrix, ($\partial u^{p'} / \partial u^p = \partial (r,\theta,\phi) / \partial (z,x,y)$), is the inverse of the $D$ matrix. Therefore the physical elements in the local Cartesian system are identical to those in the spherical polar coordinates}\deleted{such that the first-order magnetic tensor elements in the local Cartesian system by eq.\eqref{eq:nabla_caseB} are given by} given by the relations

\begin{equation}
V\added{^*}_{;z} = V\added{^*}_{\added{,}r}\quad, \quad V\added{^*}_{;x} = V\added{^*}_{\added{,}\theta}\quad, \quad V\added{^*}_{;y} = V\added{^*}_{\added{,}\phi}  \; .
\end{equation}

\added{In addition, from eqs.\eqref{eq:nabla_v_phys} and \eqref{eq:scale_spherical}, and using the fact that eq.\eqref{eq:scale_cartesian} also holds for the local Cartesian system having ($h_z=h_x=h_y=1$), we obtain the relations}

\begin{equation}
\added{V_{,z} = V_{,r} \; , \quad V_{,x} = \frac{1}{r}V_{,\theta} \; , \quad V_{,y} = \frac{1}{r\mathrm{sin}\,\theta}V_{,\phi}  \; .}
\end{equation}

The second-order tensor in the spherical polar coordinates is given by eq.\eqref{eq:nabla_nabla_V} using the Christoffel's symbols from eq.\eqref{eq:christoffel_spherical} and metric scale factors eq.\eqref{eq:scale_spherical}

\begin{align}
\nabla \nabla V &= V_{\added{,}rr} \mathbf{\hat{e}_r} \mathbf{\hat{e}_r} + \left(\frac{1}{r}  V_{\added{,}\theta r} - \frac{1}{r^2} V_{\added{,}\theta} \right) \mathbf{\hat{e}_r} \mathbf{\hat{e}_{\theta}}
+ \left( \frac{1}{r \mathrm{sin}\theta} V_{\added{,}\phi r} - \frac{1}{r^2 \mathrm{sin}\theta} V_{\added{,}\phi} \right) \mathbf{\hat{e}_r} \mathbf{\hat{e}_{\phi}} \nonumber \\
                &+ \left( \frac{1}{r} V_{\added{,}r \theta} - \frac{1}{r^2} V_{\added{,}\theta} \right) \mathbf{\hat{e}_{\theta}} \mathbf{\hat{e}_r} + \left( \frac{1}{r^2} V_{\added{,}\theta \theta} \replaced{+}{-} \frac{1}{r} V_{\added{,}r} \right) \mathbf{\hat{e}_{\theta}} \mathbf{\hat{e}_{\theta}} + \left( \frac{1}{r^2 \mathrm{sin}\theta} V_{\added{,}\phi \theta} - \frac{\mathrm{cos}\theta}{r^2 \mathrm{sin}^2\theta} V_{\added{,}\phi} \right) \mathbf{\hat{e}_{\theta}} \mathbf{\hat{e}_{\phi}}  \nonumber \\
                &+\left( \frac{1}{r \mathrm{sin}\theta} V_{\added{,}r \phi} - \frac{1}{r^2 \mathrm{sin}\theta} V_{\added{,}\phi} \right) \mathbf{\hat{e}_{\phi}} \mathbf{\hat{e}_r} + \left( \frac{1}{r^2 \mathrm{sin}\theta} V_{\added{,} \theta \phi} - \frac{\mathrm{cos}\theta}{r^2 \mathrm{sin}^2\theta} V_{\added{,}\phi} \right) \mathbf{\hat{e}_{\phi}} \mathbf{\hat{e}_{\theta}} + \cdots  \nonumber \\
                & + \left( \frac{1}{r^2 \mathrm{sin}^2\theta} V_{\added{,} \phi \phi} + \frac{1}{r} V_{\added{,}r} +  \frac{\mathrm{cos}\theta}{r^2 \mathrm{sin}\theta} V_{\added{,}\theta} \right) \mathbf{\hat{e}_{\phi}} \mathbf{\hat{e}_{\phi}} \; . \label{eq:second_order_pot}
\end{align}

\deleted{Note here the convention of notation $V_{rr}=\partial^2 V / \partial r^2$ and $V_{\theta r} = \partial^2 V / \partial \theta \partial r$ which is different from the tensor element notation i.e. $V_{;rr}$ and $V_{;\theta r}$.} 
At the position $P$ (being the GVO target point), we \added{can relate the physical components of eq.\eqref{eq:second_order_pot} to the local Cartesian system by using eq.\eqref{eq:nabla_nabla_Trans2_2} having the transformation matrix equal to the identity matrix, i.e. $R=I$. This leads to the following relations}\deleted{therefore have the following identifications between the tensor elements in the local Cartesian and the spherical systems} 

\begin{align}
V\added{^*}_{;zz}&=V\added{^*}_{;rr} \quad V\added{^*}_{;zx}=V\added{^*}_{;r \theta} \quad V\added{^*}_{;zy}=V\added{^*}_{;r \phi} \\
V\added{^*}_{;xz}&=V\added{^*}_{;\theta r} \quad V\added{^*}_{;xx}=V\added{^*}_{;\theta \theta} \quad V\added{^*}_{;xy}=V\added{^*}_{;\theta \phi} \\
V\added{^*}_{;yz}&=V\added{^*}_{;\phi r} \quad V\added{^*}_{;yx}=V\added{^*}_{;\phi \theta} \quad V\added{^*}_{;yy}=V\added{^*}_{;\phi \phi} \; .
\end{align}

\added{Since the natural and physical component are identical in the local Cartesian system (in fact in any Cartesian system), we have the relations}

\begin{align}
V_{,zz}&=V^*_{;rr} \quad V_{,zx}=V^*_{;r \theta} \quad V_{,zy}=V^*_{;r \phi} \label{eq:second_order_relat1} \\
V_{,xz}&=V^*_{;\theta r} \quad V_{,xx}=V^*_{;\theta \theta} \quad V{,xy}=V^*_{;\theta \phi} \\
V_{,yz}&=V^*_{;\phi r} \quad V_{,yx}=V^*_{;\phi \theta} \quad V_{,yy}=V^*_{;\phi \phi} \; .  \label{eq:second_order_relat3}
\end{align}

\added{By using eqs.\eqref{eq:second_order_relat1} to \eqref{eq:second_order_relat3} together with eq.\eqref{eq:second_order_pot}, we can infer the relations between the gradient tensor elements in the local Cartesian system and the gradient tensor described in the spherical system. We can do this by recognizing, that the components \deleted{(e.g. $V_{,rr}$) }in front of the elementary unit tensor (e.g. $\hat{\mathbf{e}}_r\hat{\mathbf{e}}_r$) of eq.\eqref{eq:second_order_pot}, are the physical components of the $\nabla \nabla V$ tensor, so that:}\deleted{This means that the relations between the gradient tensor elements in the local Cartesian system and the gradient tensor described in the spherical system are given by eq.\eqref{eq:nabla_nabla_caseB}}

\begin{align}
V_{\replaced{,}{;}zz} &= V_{\added{,}rr} \nonumber \\
V_{\replaced{,}{;}xz} &= \frac{1}{r} V_{\added{,}\theta r} -\frac{1}{r^2} V_{\added{,}\theta} \nonumber \\
V_{\replaced{,}{;}yz} &= \frac{1}{r\mathrm{sin}\theta} V_{\added{,}\phi r} - \frac{1}{r^2\mathrm{sin}^2\theta}V_{\added{,}\phi} \nonumber \\
V_{\replaced{,}{;}zx} &= \frac{1}{r} V_{\added{,}r \theta}-\frac{1}{r^2} V_{\added{,}\theta} \nonumber \\
V_{\replaced{,}{;}xx} &= \frac{1}{r^2} V_{\added{,}\theta \theta} +\frac{1}{r} V_{\added{,}r} \nonumber \\
V_{\replaced{,}{;}yx} &= \frac{1}{r^2\mathrm{sin}^2\theta}  V_{\added{,}\phi \theta} -\frac{\mathrm{cos}\theta}{r^2\mathrm{sin}^2\theta} V_{\added{,}\phi} \nonumber \\
V_{\replaced{,}{;}zy} &= \frac{1}{r\mathrm{sin}\theta} V_{\added{,}r \phi} - \frac{1}{r^2\mathrm{sin}\theta} V_{\added{,}\phi} \nonumber \\
V_{\replaced{,}{;}xy} &= \frac{1}{r^2\mathrm{sin}\theta} V_{\added{,}\theta \phi} - \frac{\mathrm{cos}\theta}{r^2\mathrm{sin}^2\theta} V_{\added{,}\phi} \nonumber \\
V_{\replaced{,}{;}yy} &= \frac{1}{r^2\mathrm{sin}^2\theta} V_{\added{,}\phi \phi} +\frac{1}{r} V_{\added{,}r} +\frac{\mathrm{cos}\theta}{r^2\mathrm{sin}\theta} V_{\added{,}\theta} \; . \label{eq:7}
\end{align}

In order to express the gradient tensor in Cartesian coordinates, we note that the metric tensor becomes the identity matrix meaning that the metric scale factors $h_p$ becomes unity, and all of the Christoffel's symbols becomes zero such that the gradient tensor is given by eq.\eqref{eq:nabla_nabla_V}

\begin{align}
\nabla \mathbf{B} &=  - \left(
\begin{array}{lll}
\frac{\partial^2 V}{\partial z^2}          & \frac{\partial^2 V}{\partial x \partial z} & \frac{\partial^2 V}{\partial y \partial z} \\
\frac{\partial^2 V}{\partial z \partial x} & \frac{\partial^2 V}{\partial x^2}          & \frac{\partial^2 V}{\partial y \partial x} \\
\frac{\partial^2 V}{\partial z \partial y} & \frac{\partial^2 V}{\partial x \partial y} & \frac{\partial^2 V}{\partial y^2}
\end{array}\right)= 
- \left(
\begin{array}{lll}
V_{\replaced{,}{;}zz}          & V_{\replaced{,}{;}zx}  & V_{\replaced{,}{;}zy} \\
V_{\replaced{,}{;}xz}          & V_{\replaced{,}{;}xx}  & V_{\replaced{,}{;}xy} \\
V_{\replaced{,}{;}yz}          & V_{\replaced{,}{;}yx}  & V_{\replaced{,}{;}yy}
\end{array}\right) =
\left(
\begin{array}{lll}
\left[\nabla B\right]_{zz}          & \left[\nabla B\right]_{zx}  & \left[\nabla B\right]_{zy} \\
\left[\nabla B\right]_{xz}          & \left[\nabla B\right]_{xx}  & \left[\nabla B\right]_{xy} \\
\left[\nabla B\right]_{yz}          & \left[\nabla B\right]_{yx}  & \left[\nabla B\right]_{yy}
\end{array}\right) \;,  \label{eq:tensor_cartesian}
\end{align}

where the minus sign comes from defining the field as the negative gradient of the potential. We recall that the semicolon notation denotes tensor elements following \cite{Casotto_Fantino_2009}, and not the second order spatial derivatives. However, in the case of the gradient tensor in Cartesian coordinates, these two are equivalent cf. eq.\eqref{eq:tensor_cartesian}.   

\end{document}